\documentclass[epj,twocolumn]{webofc}
\usepackage[varg]{txfonts}
\usepackage{graphicx,amsmath}

\newcommand{\vv}{\mathbf{v}}

\newcommand{\vB}{\mathbf{B}}

\newcommand{\vJ}{\mathbf{J}}

\newcommand{\bomega}{\boldsymbol{\Omega}}

\newcommand{\grad}{\boldsymbol{\nabla}}
\newcommand{\bcdot}{\boldsymbol{\cdot}}
\newcommand{\dvg}{\boldsymbol{\nabla}\!\bcdot\!}
\newcommand{\curl}{\boldsymbol{\nabla}\!\boldsymbol{\times}\!}

\newcommand{\cross}{\!\boldsymbol{\times}\!}

\wocname{epj}
\woctitle{Seismology of the Sun and the Distant Stars 2016}
\begin{document}

\title{Dynamos and Differential Rotation:}
\subtitle{Advances at the Crossroads of Analytics, Numerics, and Observations}
\author{\firstname{Kyle} \lastname{Augustson}\inst{1}\fnsep\thanks{\email{kyle.augustson@cea.fr}}}
\institute{CEA/DRF/IRFU Service d'Astrophysique, CEA-Saclay, 91191 Gif-sur-Yvette Cedex, France}

%-----------------------------------------------------------------------
\abstract{The recent observational, theoretical, and numerical progress made in understanding stellar magnetism is
  discussed. Particularly, this review will cover the physical processes thought to be at the origin of these magnetic
  fields and their variability, namely dynamo action arising from the interaction between convection, rotation,
  radiation and magnetic fields. Some care will be taken to cover recent analytical advances regarding the dynamics and
  magnetism of radiative interiors, including some thoughts on the role of a tachocline. Moreover, recent and rapidly
  advancing numerical modeling of convective dynamos will be discussed, looking at rapidly rotating convective systems,
  grand minima and scaling laws for magnetic field strength. These topics are linked to observations or their
  observational implications.}

\maketitle

%-----------------------------------------------------------------------
\section{Introduction} \label{intro}

Whence came the crossroads? With HMI's new additions to helioseismology, Kepler's magnificent
asteroseismic capabilities, and numerical breakthroughs with dynamo simulations, the last five years
have been an exciting time during which the community has uncovered a few new plot twists in the
mystery that is the dynamics occurring within stars.

\begin{figure}[!t]
\centering
\includegraphics[width=\hsize]{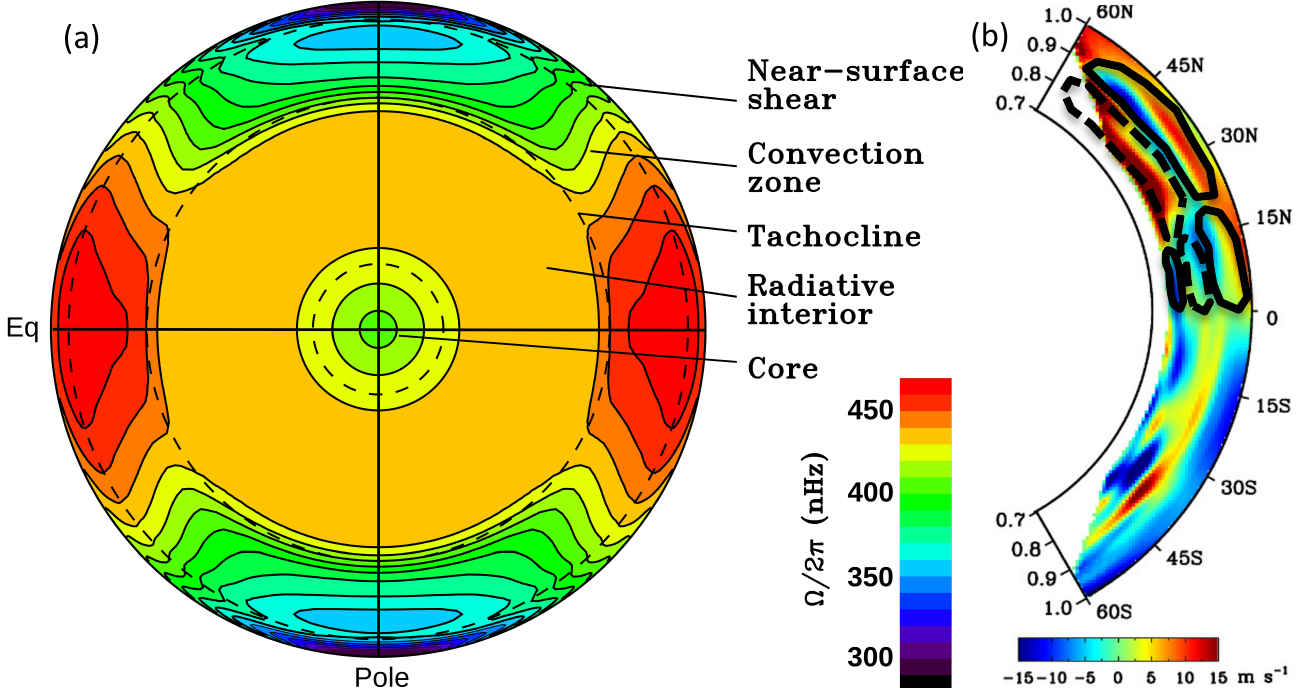}
\caption{Helioseismically-determined large-scale flows in the Sun.  (a) The solar differential
  rotation, with specific features highlighted \citep[Adapted from][]{howe09}. (b) The meridional
  circulation, with the major circulation cells highlighted.  Solid lines denote a counter-clockwise
  flow, whereas dashed lines denote a clockwise flow \citep[Adapted from][]{zhao13}.}
\label{fig:solarflows}\vspace{-0.25truein} 
\end{figure}

One such mystery that has long dogged astrophysicists has been the source of the Sun's magnetism and its variability.
While pieces of this mystery appear to be solved, there are still many vexing details that are left unexplained, such as
why the cycle period is 22 years and why it varies in both duration and amplitude.  Moreover, why surface magnetic
features migrate toward the pole and equator as the cycle advances is still not clearly determined. Also, a long-running
debate within the stellar community has been what is the precise role of a tachocline, which is the region where there
likely is a distinct change in a star's rotation profile near the boundaries of its convective zones. In particular, the
term tachocline originates from observations of the Sun's differential rotation \citep{howe09}, as shown in Figure
\ref{fig:solarflows}(a), where the latitudinal differential rotation in its convective exterior is seen to transition to
a latitudinally-uniform rotation in its convectively-stable, radiation-dominated interior. Part of the debate resides in
a tachocline's role in a dynamo, being whether or not it influences the duration of the cycle period or the even-odd
cycle parity \citep{brun15}. Another piece of the debate regards the role of turbulent diffusion processes in
homogenizing the rotation profile \citep{zahn91}, as well as its apparent paradoxical ability to prevent the spread of
the differential rotation into the stable region \citep{gough98}. Some of these topics have begun to be addressed in
global-scale dynamo simulations, as well as having been scrutinized both analytically and in local-scale simulations, as
will be discussed in \S\ref{sec:lowmass}.

%++++++++++++++
%++++++++++++++

\begin{figure*}[!t]
   \begin{center}
   \includegraphics[width=\textwidth]{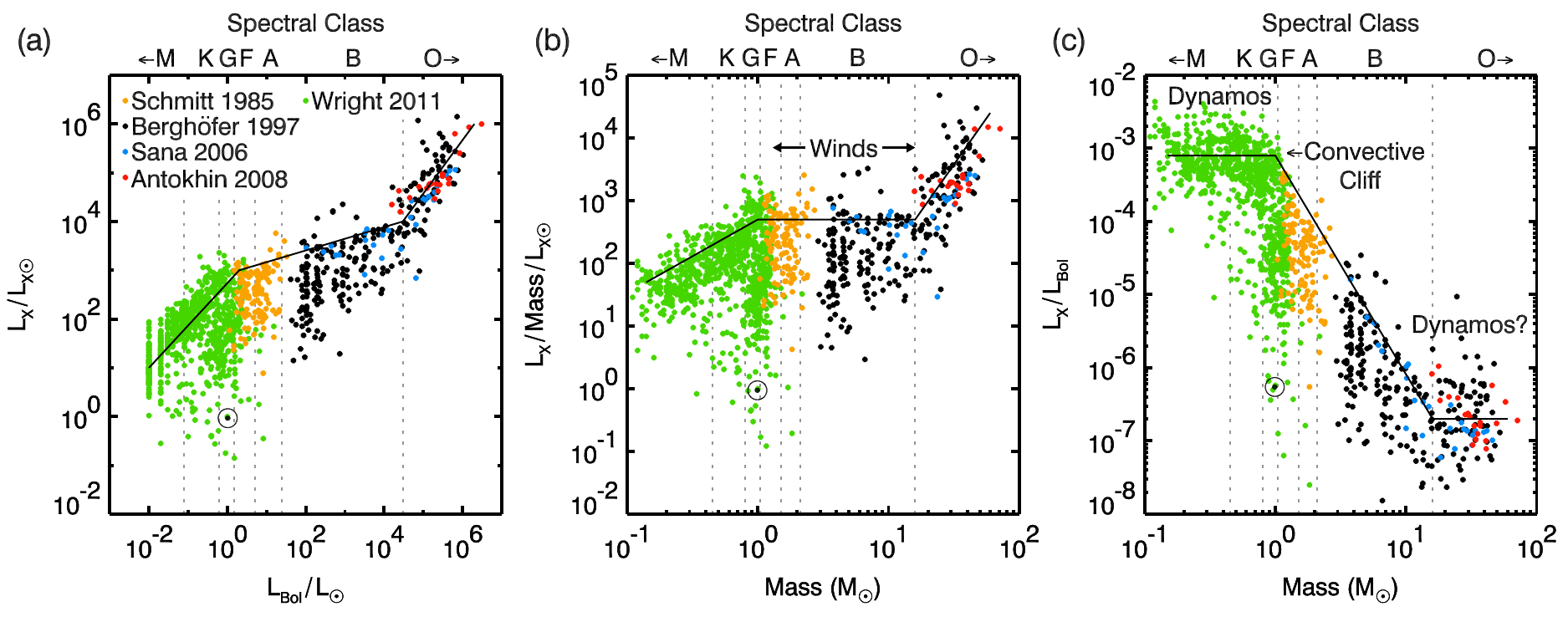}
   \caption{Stellar X-ray luminosity observations for a large range of stellar masses. (a) Average X-ray luminosity 
     $L_\mathrm{X}$ versus bolometric luminosity $L_{\mathrm{Bol}}$ normalized by their respective solar values. 
     Spectral type is indicated along the top abscissa and demarked with thin dotted lines. The black line illustrates 
     the average behavior for three distinct mass ranges.  The data sources are 
     \citep{schmitt85,berghoefer97,sana06,antokhin08,wrightn11}. (b) $L_{\mathrm{X}}$ to mass ratio plotted versus mass,
     emphasizing a mass range with wind-driven emission where $L_{\mathrm{X}}$ depends linearly on mass. (c) The ratio 
     $L_{\mathrm{X}}/L_{\mathrm{Bol}}$ versus mass, which highlights the mass range where $L_{\mathrm{X}}$ is linearly 
     related to $L_{\mathrm{Bol}}$. The location of the Sun is plotted with its symbol $\odot$, showing its relatively 
     low X-ray luminosity.}\vspace{-0.25truein}
   \label{fig:lxlbol}
   \end{center}
\end{figure*}

One such mystery was lurking in the large set of Kepler data for red giants.  When parsed in the diction of
asteroseismology, this data yields the measurement of many mixed modes.  These modes behave like an acoustic wave
(p-mode) near the stellar surface, but change their character to buoyancy-driven gravity waves (g-modes) within the
convectively-stable core of those red giants \citep[e.g.,][]{dalsgaard04,dupret09}, where these convectively-excited
modes evanescently tunnel into the core. These modes can thus provide both information about the envelope and the core
of the red giant, such as identifying the current evolutionary state of the star \citep[e.g.,][]{bedding11, mosser14},
as well as assessing the relative differential rotation of these two zones \citep[e.g.,][]{beck12,deheuvels15}. The
enigma hidden in this data was the question: why are the dipole mixed modes suppressed for a some red giants and not for
others? An astute prediction proposes that this effect is due to the magnetic field present within and near the core of
red giants, being dubbed the magnetic ``greenhouse effect'' \citep{fuller15,cantiello16}. Thus, Kepler observations have
revealed the asteroseismic traces of the remnants of potentially very strong magnetic fields that were earlier built
deep within the convective core of intermediate mass stars, such as the F-type and A-type stars.  Such dynamos will be
the topic of \S\ref{sec:massive}.

\section{Magnetism Across the HR-Diagram}\label{sec:maghr}

The physics underlying the generation of magnetic fields in astrophysical objects, including stars, appears to be quite
generic, with an interaction of plasma motion and large-scale rotation playing key roles. In stars, the most vigorous
motions occur in their convective regions.  Massive stars have a convective core below a radiative exterior, whereas
stars less massive than about 1.6~$M_{\odot}$ have an outer convective envelope above a radiative zone, and the lowest
mass stars are fully convective. The magnetic field generated in those regions is largely responsible for the majority
of the observed magnetic phenomena in, or above, the stellar photosphere such as starspots, a chromosphere, coronal
plasma, flares, and coronal mass ejections. The detailed study of our nearest star, the Sun, has provided at least a
framework for interpreting the X-ray emissions of other cool stars \citep{peres00}. Indeed, most cool stars, except cool
giant stars, have been found to exhibit characteristics grossly similar to what we know from the Sun. Most of these stars
have variable X-ray emission at temperatures of at least $1$-$2$~MK and occasional flaring.

Some of the behavior exhibited by low, intermediate, and high mass stars is displayed in Figure \ref{fig:lxlbol}. There
appear to be three distinct mass (or luminosity) ranges where there are changes in the power law relationship between
X-ray luminosity ($L_X$) and bolometric luminosity as in Figure \ref{fig:lxlbol}(a). The $L_X$-Mass ratio in Figure
\ref{fig:lxlbol}(b) shows that the intermediate mass range between the low-mass A-type stars and the higher-mass B-type
stars has $L_{\mathrm{X}} \propto M$, which may be due to strong shocks in their winds.  When scaled by
$L_{\mathrm{Bol}}$, as in Figure \ref{fig:lxlbol}(c), two other regimes become evident, with one for the lower mass
stars and another for the massive O-type stars where $L_{\mathrm{X}} \propto L_{\mathrm{Bol}}$.  The lower mass stars
have a relatively high level of X-ray emission, whereas the relative X-ray emission declines significantly with higher
mass and becomes nearly constant beyond about 16~$M_{\odot}$.

The precise nature of the physical processes driving these changes in X-ray emission is not known. Though, as always,
there are a few hints about what might be occurring. For instance, the saturation of $L_{\mathrm{X}}$ at low masses is
potentially a result of the strong magnetic confinement of their winds or a saturation of those stars dynamos, both of
which could be due to the rapid rotation of the star. The saturation for higher mass stars may result from two
mechanisms: one could be dynamo action in the Fe- and He-opacity bump convection zones; the other may be due to changing
properties of the winds and mass loss for the O-type stars. It may be possible to disentangle these processes through
asteroseismic measurements of these higher-mass stars, looking for a shift from a dominance of g-modes to p-modes.

\section{Magnetic Activity-Rotation Correlation} \label{sec:magact}

\begin{figure}[t!]
   \begin{center}
   \includegraphics[width=0.8\hsize]{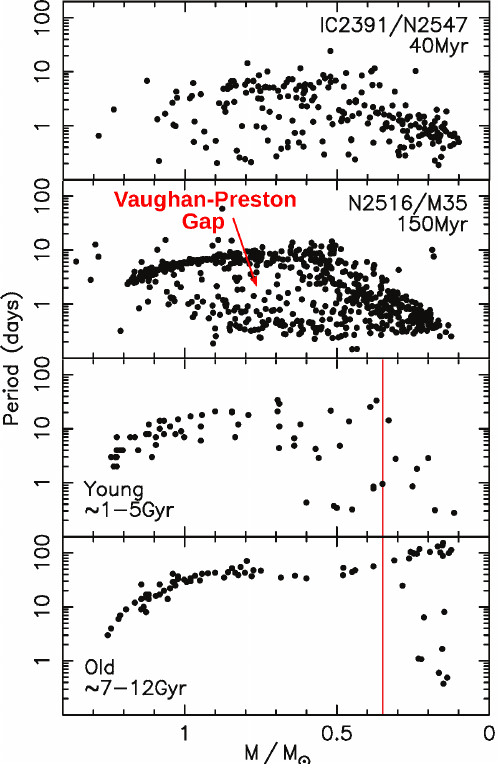}
   \caption{Rotation period distributions of stellar populations of differing ages and mass, showing the initially
     uniform distribution of a very young 1~Myr old cluster that evolves toward slower rotation with greater age. The
     panel in the upper right illustrates the Vaughan-Preston Gap, with a population of slow rotators, of fast rotators,
     and very few stars between them. The red line in the lower right two panels denotes the transition at
     0.3~$M_{\odot}$ from stars with a radiative interior to those that are fully convective (Adapted from
     \citep{stassun11}).} \label{fig:periodage}\vspace{-0.25truein}
   \end{center}
\end{figure}

William Herschel was the first to extend the contemporary idea of the rotation of the Sun and the planets to other stars
\citep{herschel1795}. In fact, he went so far as to attribute the periodic light curves that he observed for other stars
to star spots rotating in and out of the line-of-sight \citep{herschel1796}. Eighty years later, the first Doppler
measurements of the Sun \citep{abney1877a} were made confirming its rotation. Shortly thereafter and using similar
techniques, studies of stellar rotation shed light on two regimes of stellar rotation along the main-sequence that
depends upon mass \citep[e.g.,][]{shajn29,wilson66}. In particular, they found that stars more massive than about
1.6~$M_{\odot}$ rotate fairly rapidly at an average speed of around 140~$\mathrm{km s^{-1}}$ with a slow increase with
mass and reaching a plateau above about 10~$M_{\odot}$. Stars less massive than 1.6~$M_{\odot}$, on the other hand, show
a steep decline in $v\sin{i}$ with decreasing mass.  There is a distribution of their projected equatorial rotation
speed ($v\sin{i}$) about this mean, as the statistics in the above references show and as can be seen in Figure
\ref{fig:periodage}.

Spectroscopic surveys of nearby field stars and clusters reveal a potential reason for the rapid decline of a star's
rotational period with age. Based on observations of chromospheric lines in young clusters and old field stars
\citep[e.g.,][]{wilson63}, it is apparent that magnetic activity declines with age. Indeed, the observational link between
rotation, convection, and magnetic activity was described in \citep{wilson66}, where an attempt is made to ascertain why
stars less massive than about 1.5~$M_{\odot}$ rotate more slowly than more massive stars. A direct link between these
physical processes is made when employing concepts pioneered in dynamo theory \citep{moffatt78, brun15}. As such, it is
posited that the loss of an outer convection zone leads to an inability to generate a strong, large-scale magnetic field
that would otherwise efficiently brake a star's rotation. Though, this logic would not hold for the most massive stars
as they too have outer convection zones, however their relatively short lives lead to less time for wind-related
breaking on the main-sequence even if their mass-loss rate is greater.

For lower-mass stars, the link between age, magnetic activity, and rotation rate was put into perspective and formulated
into what is now widely accepted as the ``Skumanich $t^{-1/2}$ law'' for the decay of a star's rotation rate and
chromospheric magnetic activity \citep{skumanich72}. In particular, two regimes of dynamo behavior seem apparent along
the main sequence. As the star spins down, there is a transition between two regimes of chromospheric activity, with a
sparsely populated gap between them dubbed the ``Vaughan-Preston Gap'' \citep[e.g.,][]{vaughan80}. This can be seen most
easily in Figure \ref{fig:periodage}, which shows the rotation period versus mass for several clusters of increasing
age.

\begin{figure}[t!]
   \begin{center}
   \includegraphics[width=\hsize]{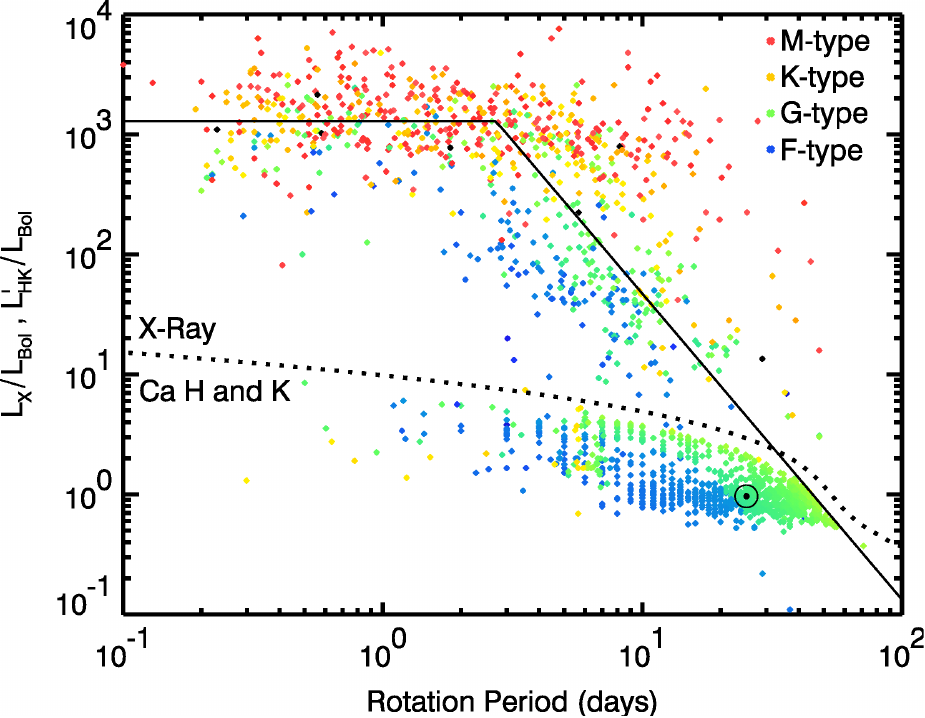}
   \caption{X-ray and Ca II observations of individual stars are shown, which cover a range of rotation rates and masses
     for lower mass stars. Ratios of X-ray luminosity $L_{\mathrm{X}}$ and chromospheric Ca II luminosity
     ($L'_{\mathrm{HK}}$) to the bolometric value $L_{\mathrm{Bol}}$ (normalized so that these ratios are unity for the
     Sun; $L_{\mathrm{X}}/L_{\mathrm{Bol}}\approx10^{-6}$ and $L'_{\mathrm{HK}}/L_{\mathrm{Bol}}\approx10^{-5}$),
     displayed with rotational periods as abscissa. The color indicates a star's mass, with M-type in red, K-type in
     yellow, G-type in green, and F-type in blue. The dotted line demarks the boundary between X-ray and Ca II
     measurements. The solid line shows an estimated fit for the Rossby number scaling of $L_X$.  The location of the
     Sun is shown with its symbol ($\odot$). Constructed with Ca II data from \citep{soderblom93a,wrightj04} and X-ray
     data from \citep{wrightn11}.} \label{fig:actrot}\vspace{-0.25truein}
   \end{center}
\end{figure}

The relationship between stellar rotation and proxies of magnetic activity remains an important diagnostic of the dynamo
processes likely responsible for stellar magnetism in low mass stars.  Chromospheric variability measured in Ca II, and
other chromospheric lines, as well as the X-ray observations that probe stellar coronae, indicate that there is a
monotonic increase in activity with rotation rate. This increase is halted as the rotation rate reaches a critical
point, after which either these indicators of magnetic activity or these star's dynamos become saturated. Indeed, both
coronal and chromospheric activity measures show such a saturation (Figure \ref{fig:actrot}), though it is more
prominent in the X-ray observations.

The first tentative connection between X-ray luminosity and rotation was established by \citep{pallavicini81}, reporting
that X-ray luminosity scaled as $L_X\propto (v \sin{i})^{1.9}$ for low mass stars and that the X-ray luminosities of
higher mass stars depends only on the bolometric luminosity. The scaling with rotation for low mass stars suggested
that, like the Sun, X-ray emitting coronal magnetic activity is linked to the rotationally-constrained convection that
drives the magnetic dynamo running in the outer convection zones of these stars, whereas more massive stars without such
an outer convection zone likely produce X-rays primarily through shocks in their powerful winds. With more data, the
nearly quadratic proportionality between $L_X$ and $v\sin{i}$ was found to have a limited domain of applicability, since
the ratio of X-ray to bolometric luminosity reaches a peak saturation of about $10^{-3}$ that is effectively independent
of rotation rate and spectral type \citep[e.g.,][]{vilhu84,micela85,vilhu87}. The rotational period at which this
saturation level is reached, however, does depend upon spectral type, with lower mass stars reaching this state at
longer periods when compared to higher mass stars (Figure \ref{fig:actrot}a). This can be attributed to the decrease in
the vigor of the convection as one moves down the range of masses \citep{pizzolato03}.

It is unclear whether the constancy of the X-ray luminosity below a threshold rotation period is
derived of a fundamental limit on the dynamo itself \citep[e.g.][]{vilhu84}, a quenching arising
from the pervasion of strong magnetic structures across much of the stellar surface \citep{vilhu84},
or a centrifugal stripping of the corona caused by the high rotation rates \citep{jardine99}. There
could also be changes in the X-ray production mechanisms as well as their X-ray spectral
content. However, recent work has aimed at modelling the coupling of a simple dynamo to a wind
model, with the aim of describing this transition through fairly simple physics
\citep{blackman15,blackman16}.

Characteristics similar to the X-ray flux also hold for chromospheric measures of magnetic activity
as can be seen in Figure \ref{fig:actrot}.  This figure shows the ratio of chromospheric Ca H and K
luminosity $L'_{\mathrm{HK}}$ to bolometric luminosity for a sample of main-sequence stars ranging
in mass between 0.3 and 1.6~$M_{\odot}$, with data from \citep{soderblom93a} and
\citep{wrightj04}. The ratio of X-ray luminosity and bolometric luminosity, also plotted with data
from \citep{wrightn11}, has a much greater dynamic range, spanning six orders of magnitude compared
to the roughly 1.5 orders of magnitude for the Ca H and K observations. Yet both show an increase of
activity with decreasing rotation period and an eventual saturation. It is of note that the
chromospheric activity seems to saturate at longer periods and that there are few stars that are
rotating rapidly that show Ca H and K emission, at least in this sample.

\begin{figure}[t!]
   \begin{center}
   \includegraphics[width=0.8\hsize]{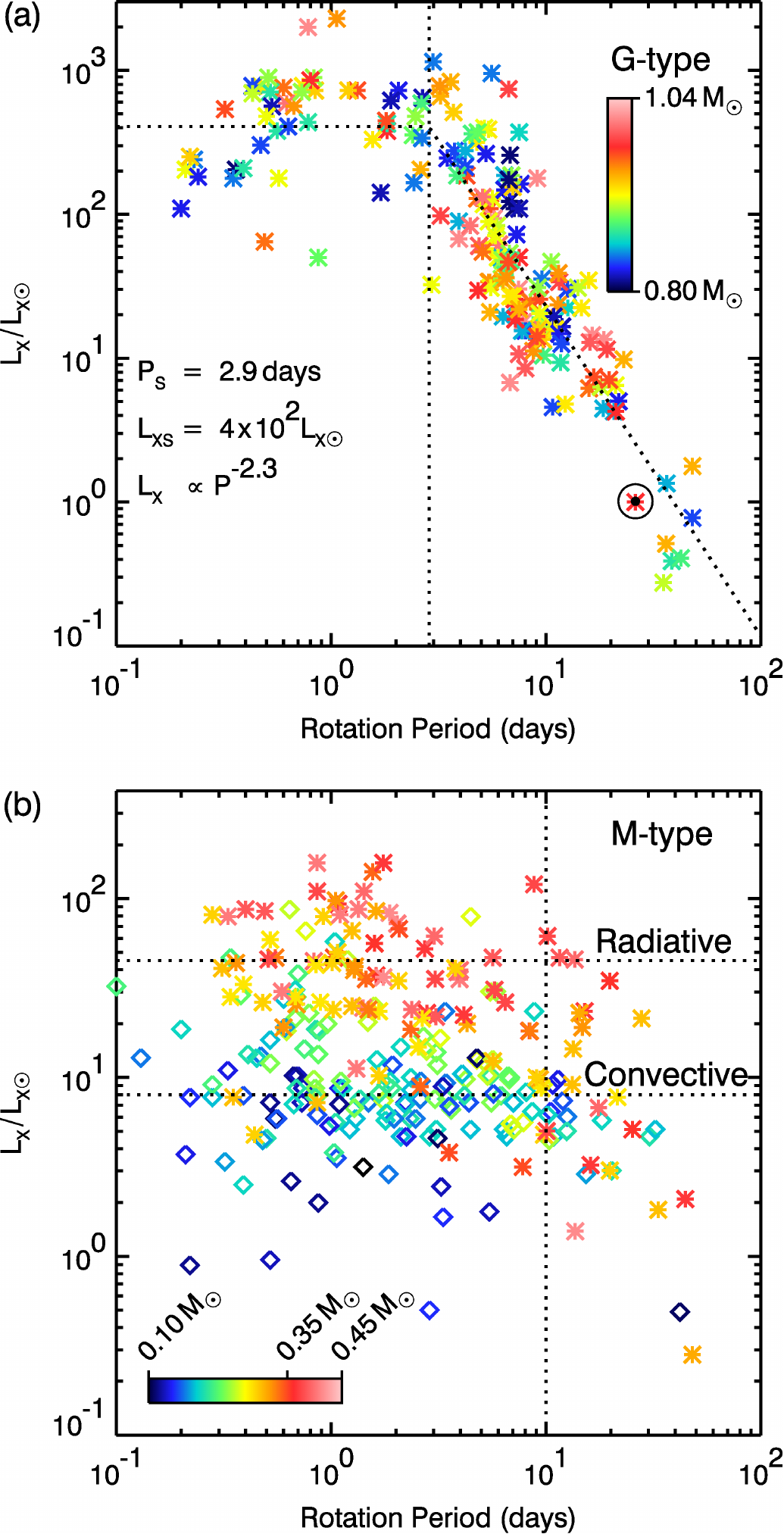}
   \caption{X-ray observations are shown for individual stars that are binned by their spectral
     type, which cover a range of rotation periods and masses. X-ray luminosity ($L_X$) normalized
     by the solar value and displayed with rotation period in days as abscissa. The color indicates
     a star's mass within the spectral type. (a) Spectral type G stars, with the Sun's position
     indicated. (b) Spectral type M stars, with stars above 0.3~$M_{\odot}$ as starred symbols and
     those below this mass as diamonds. The dotted lines indicate fits to the data for the scaling
     with period, the saturation value, and the rotation period corresponding to the beginning of
     the saturation (Constructed with data from \citep{wrightn11}).} \label{fig:lx_spec}\vspace{-0.25truein}
   \end{center}
\end{figure}

If we now separate the X-ray observations by spectral type, the trends become more obvious as in
Figure \ref{fig:lx_spec}(a) for the G-type stars. Speaking broadly, the low mass stars of K, G, and
F-type share similar characteristics in their X-ray emission involving increasing emission with more
rapid rotation and a regime of saturated emission. What differs between them is the rapidity of the
onset of this state of constant emission with rotation, the rotation period at which it saturates,
and the maximum luminosity. The maximum luminosity seems to be determined by the bolometric
luminosity, as surmised from Figure \ref{fig:actrot} where the data collapse onto the same fit. The
rotation period at which it saturates may be related to the maximum achievable dynamo action,
wherein the dynamo can no longer extract energy from the star's differential rotation due to strong
Lorentz feedbacks.

The rate of increase of X-ray emission with decreasing rotation period suggests a direct link to
the Rossby number of the flows within these stars in that it measures the efficiency of the
interaction of rotation and convection in building both the differential rotation and the magnetic
field. In particular, there appears to be a change in the slope of the decrease of X-ray emission
with rotation period, from 1.7 to 2.5 between the F and K-type stars. This may indicate that the
K-type stars have the most efficient conversion of kinetic to magnetic energy. An interesting
feature of stars about the mass of the Sun, namely the G and F-type stars, is that there seems to be
a regime of super-saturation with the X-ray luminosity decreasing at the rates of greatest rotation,
which is a topic of ongoing research. What is also evident from all of these observations is that
the Sun, while thought of as a magnetically active and robustly influential object, is in fact
hundreds of times less active in both chromospheric and coronal proxies than some of its more
rapidly rotating brethren, which serves to accentuate the overwhelming activity that must be present
at the surfaces of these other stars.

\section{Tiny Stars with Strong Magnetic Fields, and the Influence of a Tachocline} \label{sec:lowmass}

%++++++++++++++
\begin{figure}[!t]
\centering
\includegraphics[width=\hsize]{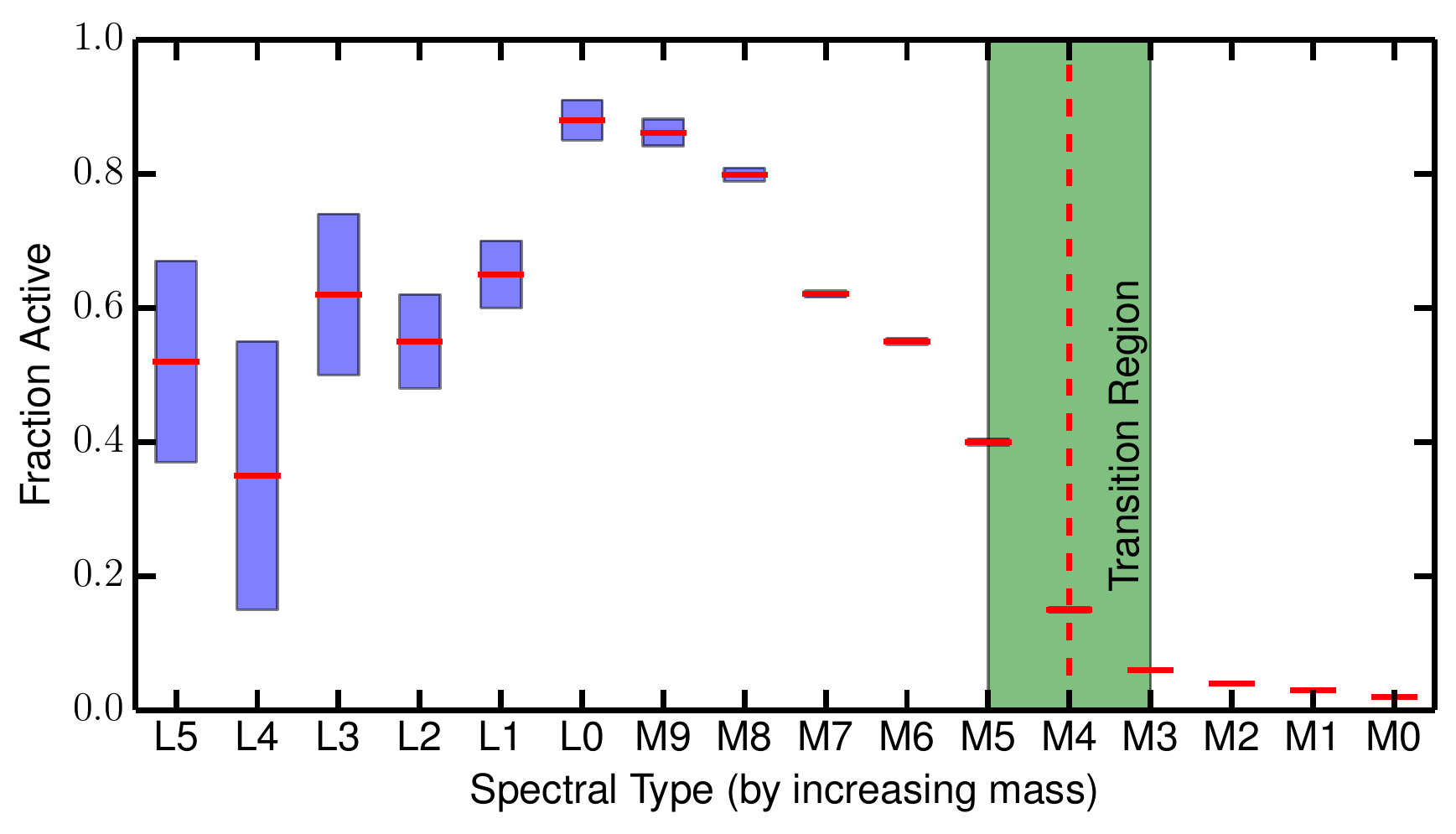}
\caption{Fraction of active L and M dwarfs, showing the transition from a low to high activity for
  stars less massive than spectral type M4 \citep[Adapted from data in][]{schmidt15}. The red line
  denotes the mean value and the blue boxes show the range of uncertainty.}\vspace{-0.25truein}
\label{fig:mdwarfact} 
\end{figure}
%++++++++++++++

In the Sun, the tachocline is a site of strong radial shear revealed by helioseismology , which is a
matching layer between the differentially rotating outer convection zone and an interior rotating as
a solid body (see Figure \ref{fig:solarflows}). Other stars with a radiative interior may also have
a tachocline. In an interface-type dynamo, the tachocline is thought to play a major role in the
overall dynamo action. Such a layer below the convection zone can store and, through the radial
differential rotation, comb small-scale magnetic field into a larger toroidal structure. When the
magnetic field strength is sufficiently large, this field can become buoyant and thus could provide
the seed magnetic field for the convective dynamo in the exterior convection zone
\citep{spiegel80,parker93}.

What is possibly more remarkable in the X-ray data is the transition between stars that possess a radiative core and
those that are fully convective (Figure \ref{fig:lx_spec}(b)). Indeed, most of the stars in our galaxy are less massive
than our Sun, with more than half of all stars being M-dwarfs. These are stars that are less than half the mass of the
Sun and thus between tens and thousands of times less luminous than it. These low-mass stars had been believed to harbor
magnetic dynamos quite distinct from those in Sun-like stars. It seems appropriate that there should be a difference
between stars that have an interface-type dynamo, with a region of penetration and convective overshoot, and those that
are fully convective.

An intriguing suite of observations of low-mass stars in H$\alpha$ indicates that there is a
transition in magnetic activity lifetime from the more massive stars that become relatively inactive
after about one billion years to the lower-mass fully-convective stars, which appear to remain
active for much longer \citep{west08}. Indeed, the trend likely continues past the pp-chain fusion
limit into the category of giant planets \citep{schmidt15}, with those objects potentially remaining
active for their lifetimes.  Moreover, as seen in Figure \ref{fig:mdwarfact}, the fraction of
spectral-type L and M dwarfs that are active in H$\alpha$ drops from a peak value of about 85\% at
spectral-type M9 to about 2\% at spectral type M0.

A trend that is likely related can also be seen in open cluster data, wherein the variance in the
distribution of the rotation rate for stars in the cluster decreases with time \citep{stassun11}.
For stars with masses between about 0.3 and 1.3 solar masses that are evolving along the main
sequence, the rotation-rate distribution slowly transits from a nearly uniform distribution to a
tight distribution about a mean value. However, if the stars' masses fall below the fully-convective
threshold value, the distribution of rotation rates appears to remain nearly uniform. This picture
of long-term activity remaining even if the star spins down is largely consistent with observations
both in X-ray, H$\alpha$, and Ca II spectral bands.

Hence, for these three observational measures of rotation rate, activity fraction, and activity
lifetime, the behavioral transition occurs over a specific range of stellar mass.  In particular, it
occurs where main-sequence stars shift from having a convective envelope to being fully convective.
Furthermore, such a transition may demark a region where a tachocline has a decreasing influence on
the star's magnetic and rotational evolution, given that the radius of the convectively-stable
radiative interior decreases for lower masses.

Stitching these pieces together hints at a question: it appears that stars with a tachocline, or at
least with a radiative interior, spin down more rapidly than their fully-convective counter parts,
so is the spin-down related to the internal dynamo?  When comparing these two kinds of stars, the
fully-convective stars tend to have stronger magnetic fields with simpler geometry
\citep{jardine02,vidotto14}. They are also more rapidly rotating. This all should lead to a more
fierce wind and stronger angular momentum loss \citep{reville15}. So, why is it that these stars
remain active so long and retain much of their zero-age main-sequence angular velocity? It may be
that the presence of a stable region is a critical component in the dynamo that inherently drives
these stars' magnetic activity and hence impacts their rotational history. Yet precisely how a such
a region could influence both of these facets of stellar evolution is unclear.

\section{Prandtl Numbers and Large and Small-scale Dynamos} \label{sec:prandtl}

A powerful tool currently available in modern dynamo theory is 3D numerical simulations. In direct
numerical simulations (DNS), all the turbulent processes are fully captured, from the driving scale
to the diffusion scales. Although these simulations are typically far removed from the length scales
important to astrophysical processes, they can shed light on the small-scale dynamics that strongly
influence the transport properties of the larger-scale system. The aim of these simulations is
usually to try to make contact with theoretical predictions of turbulent spectra, such as the
Kolmogorov scaling. However, as with large-eddy simulations (LES) that parameterize the small-scale
physics, they are limited by computational resources in obtaining ever larger resolution and are
constrained in their ability to run large suites of simulations that accurately probe large swaths
of parameter space.

There are many approaches to these problems: some that are true DNS in the sense that a particular
physical system with its empirically determined diffusion coefficients are simulated, some that
utilize other diffusive operators, and still others that employ only an implicit numerical
diffusion. The first method assumes that the number of atoms within a grid cell is still large
enough that the diffusion approximation to the collisional interactions is valid, whereas the last
assumes effectively the same thing but with a numerically defined diffusion coefficient. This
numerical diffusion could be considered a running coupling constant under the nomenclature of
renormalization theory. The same can be said for LES schemes that explicitly and artificially
increase the level of the Laplacian diffusion in the system. Thus there is some connection between
these two schemes, though it may not be simple due to the anistropropies that are often present in
LES of global-scale flows.

For simulations retaining the diffusion approximation to collisional interactions, there are two dynamo regimes one of
low magnetic Prandtl number ($Pm$) and another of large $Pm$. Theoretically, one should expect that at large $Pm$ the
magnetic Reynolds number is much greater than that of the fluid Reynolds number, which leads to the length scales of the
magnetic field being smaller than that of the velocity field. At low $Pm$, the expectation should be the opposite that
the generated magnetic fields roughly tend to be of larger scale than the velocity field. This does not preclude smaller
scale intermittent fields for the low-$Pm$ regime or vice versa for the converse, rather it is more a statement about
which components of the magnetic field contain the most energy: the large-scales for low-$Pm$ dynamos and the
small-scales for high-$Pm$ dynamos. Using the Braginskii definitions of the viscous and magnetic diffusivities
\citep{braginskii65} and stellar models, one can determine the ranges of magnetic Prandtl numbers within a star. For
dynamos in stars with convective exteriors, the low-$Pm$ regime is of greatest interest. Using the Sun as an example,
the molecular $Pm$ ranges between about $10^{-1}$ in the core to roughly $10^{-6}$ at the photosphere, decreasing
exponentially toward it. Other low-mass, main-sequence stars are roughly similar, with a low $Pm$ throughout the bulk of
their domains. Higher-mass, main-sequence stars, in contrast, have relatively large molecular values of $Pm$ ranging
from about $10$ throughout the convective core and remaining nearly unity until the near-surface region is reached at
which point it drops to about $10^{-2}$. So, the convective cores of massive stars and the convective envelopes of low
mass stars represent very different dynamo regimes with respect to the scales possessing the greatest magnetic energy.

\begin{figure}[t!]
  \begin{center}
    \includegraphics[width=0.8\hsize]{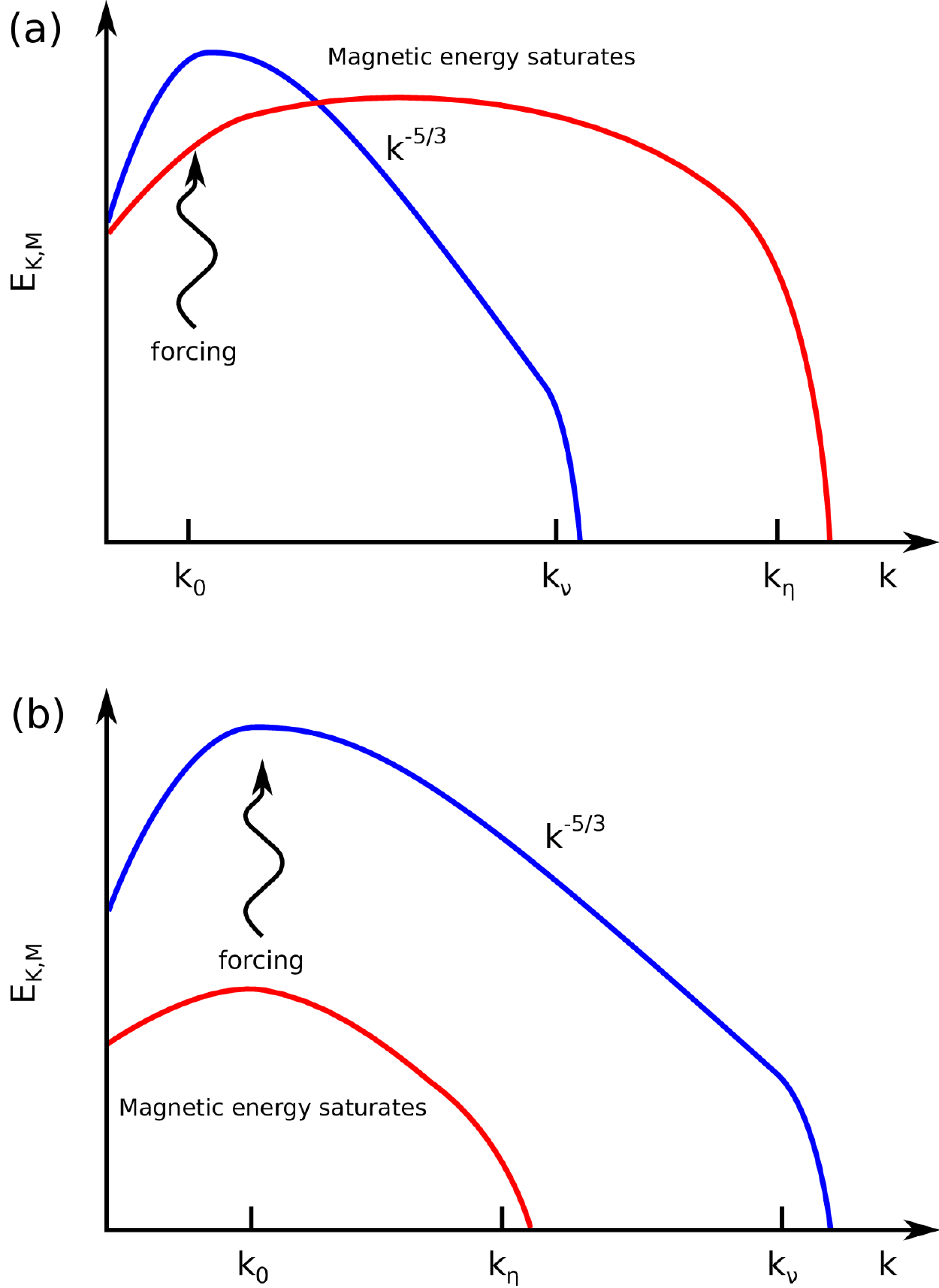}
    \caption{A sketch of energy spectra for low and high-$Pm$ dynamos with wavenumber $k$. (a)
      Saturation phase of the high-$Pm$ dynamo, with kinetic energy in blue and magnetic energy in
      red. (b) Saturation phase of the low-$Pm$ dynamo. Here $k_0$ is the driving scale, $k_{\nu}$
      the viscous dissipation scale, and $k_{\eta}$ the magnetic energy dissipation
      scale. \label{fig:dynamopm}}\vspace{-0.25truein}
    \end{center}
\end{figure}

At small scales, the $Pm$ parameter regimes have been studied extensively with numerical
simulations. The primary results of these simulations are that, at large magnetic Prandtl number,
the magnetic field generation is of the stretch-twist-fold variety where its structures are of a
smaller scale than the viscous dissipation scale \citep[e.g.,][]{schekochihin04}.  On the other side
of the Prandtl number coin, the low $Pm$ regime has the stretch-twist-reconnect mechanism, where
the flows act similar to the large $Pm$ regime, but instead of the flows dissipating and the
magnetic field folding on itself, the magnetic field dissipates leading to reconnection
\citep[e.g.,][]{schekochihin07,brandenburg09}. These two dynamo regimes are sketched in Figure
\ref{fig:dynamopm}, showing the saturated phase of the dynamo. While such simulations are
instructive for creating subgrid-scale models and understanding the small-scale physics, they lack a
direct connection to the global-scale phenomena that are more directly observable for astrophysical
objects.

Likewise, there are many well-studied analytical or semi-analytical large-scale dynamos, for instance consider those of
\citep{parker55}, \citep{backus58}, \citep{herzenberg58}, and \citep{roberts70}. These dynamos generate magnetic fields
that have spatial scales greater than the velocity scales producing them. The Sun and the Earth, among many other
astrophysical objects, seem to possess global-scale fields and as such they are likely large-scale dynamos with a
cascade of energy moving from small-scale turbulence to large-scale magnetic fields. For stellar and planetary dynamos,
the mechanisms building large-scale field is directly linked to their rotation. In these global-scale dynamos, one
manner in which the rotation is crucial is that the Coriolis force induces a net kinetic helicity in the small-scale
turbulent flows.  Correlations between the fluctuating velocity components of the turbulence consequently yield Reynolds
stresses that act to maintain the bulk differential rotation through an inverse cascade of kinetic helicity and
energy. This inverse cascade is thus the indirect source of the differential rotation that fuels the generation of
toroidal field, whereas direct action of the convection is largely to transform this toroidal field into poloidal field
as in an $\alpha$-effect.  Thus the structure of the convection and the differential rotation that it supports is key to
the overall nature of the global-scale stellar dynamo.

If the entire spectrum of the convective dynamo and its associated dynamics is resolved, the dissipation of the energy
injected into the system is governed by the molecular values of the diffusivities.  Using Braginskii plasma
diffusivities for a solar-like metallicity \citep{braginskii65} and a set of MESA stellar models to obtain the density
and temperature profiles \citep{paxton11}, one can define the average expected molecular magnetic Prandtl number in
either the convective core of a massive star or the convective envelope of a lower mass star.  To be specific, the
equations of interest are

\vspace{-0.25truein}
\begin{center}
  \begin{align}
    \nu = 3.2 10^{-5} \frac{T_e^{5/2}}{\rho \Lambda}, \\
    \kappa_{\mathrm{cond}} = 1.8 10^{-6} \frac{T_e^{5/2}}{\rho\Lambda}, \\
    \eta = 1.2 10^6 \Lambda T_e^{-3/2},
  \end{align}
\end{center}

\noindent where $\nu$ is the kinematic viscosity, $\eta$ is the magnetic diffusivity, $\kappa_{\mathrm{cond}}$ is the
electron thermal diffusivity, $\kappa_{\mathrm{rad}}$ is the radiative thermal diffusivity, $T_e$ is the electron
temperature in electron volts, and $\Lambda$ is the Coulomb logarithm.  These were computed assuming the conditions of a
MHD plasma. In particular, it is assumed that charge neutrality holds and that the temperatures are not excessively
high, so the Coulomb logarithm is well-defined. Moreover, in the charge-neutral regime, the magnetic diffusivity is
density independent because the electron collision time scales as the inverse power of ion density and the conductivity
is proportional to the electron density times the electron collision time. So, one can find that the thermal Prandtl
number $Pr$ and magnetic Prandtl number $Pm$ scale as

\vspace{-0.25truein}
\begin{center}
  \begin{align}
    Pr = \frac{\nu}{(\kappa_{\mathrm{cond}}+\kappa_{\mathrm{rad}})}, \\
    Pm = 2.7 10^{-11} \frac{T_e^{4}}{\rho \Lambda^2}. 
  \end{align}
\end{center}

\noindent In most cases, the radiative diffusivity is several orders of magnitude larger than the electron conductivity,
so it dominates the thermal Prandtl number.

\begin{figure}[t!]
  \centering
  \includegraphics[width=0.95\linewidth]{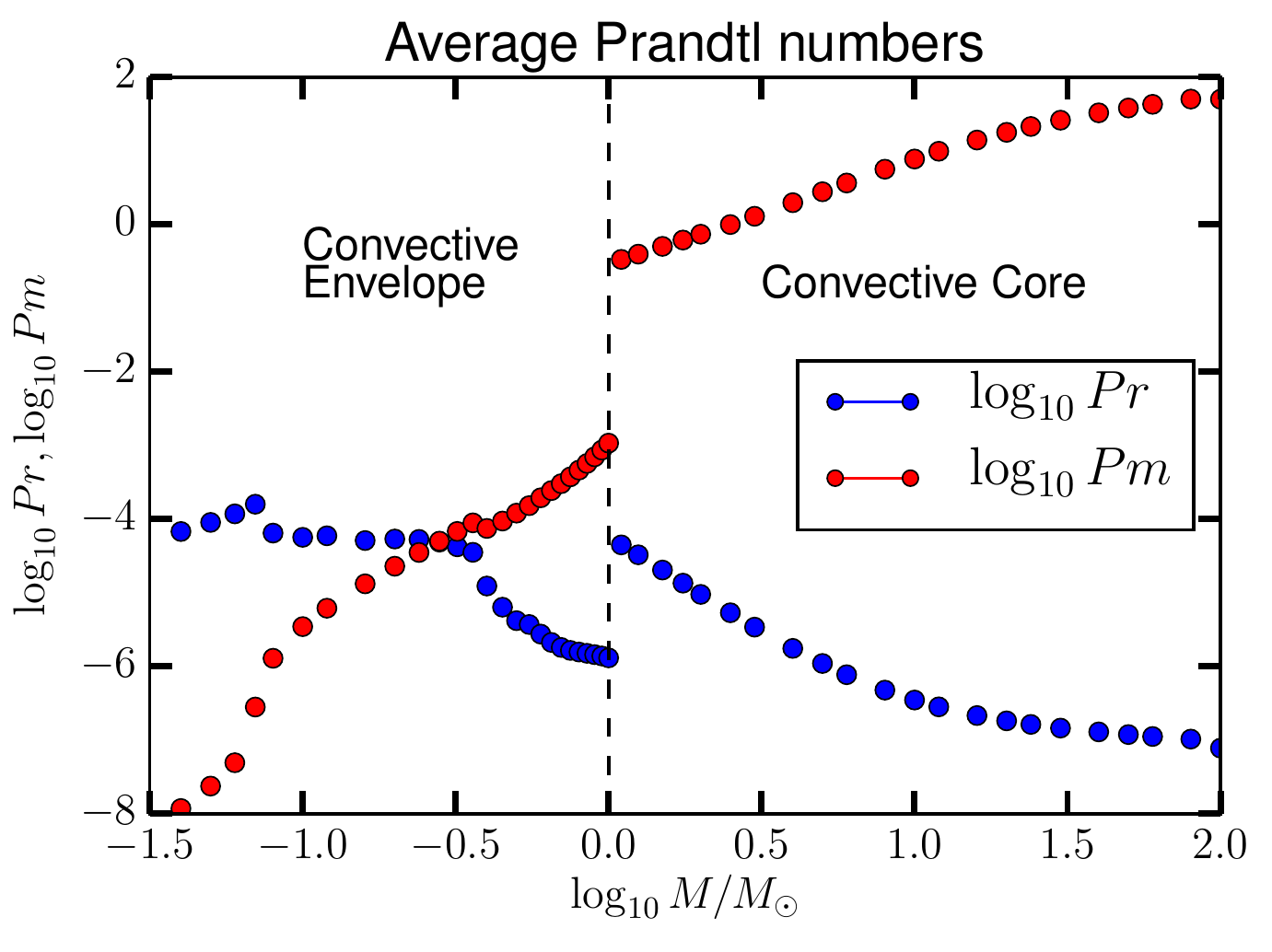}
  \caption{The thermal Prandtl number $Pr$ (blue) and magnetic Prandtl number $Pm$ (red) for stars
    with masses between 0.03 and 100~$M_{\odot}$. The convective region being averaged is the core
    for stars above about 1~$M_{\odot}$, and is either the convective envelope for stars below about
    1~$M_{\odot}$ or full star for stars below 0.3~$M_{\odot}$.}\vspace{-0.25truein}
  \label{fig:prandtl}
\end{figure}

This prescription for the Prandtl numbers has been applied to MESA models of stars on the zero-age main-sequence in the
mass range between 0.03 and 100 $M_{\odot}$ as shown in Figure (\ref{fig:prandtl}).  There, it is clear that all stars
fall into the low thermal Prandtl number regime, whereas one can find two regimes of magnetic Prandtl number. This is
directly related to where the convective region is located. For massive stars with convective cores, the temperature and
density averaged over the convective volume are quite high when compared to lower-mass stars with a convective
envelope. Such a high temperature leads to a large magnetic Prandtl number.  This implies that there could be two
fundamentally different kinds of convective dynamo action in low-mass versus high-mass stars.

\begin{center}
  \begin{figure*}[t]
    \includegraphics[width=\textwidth]{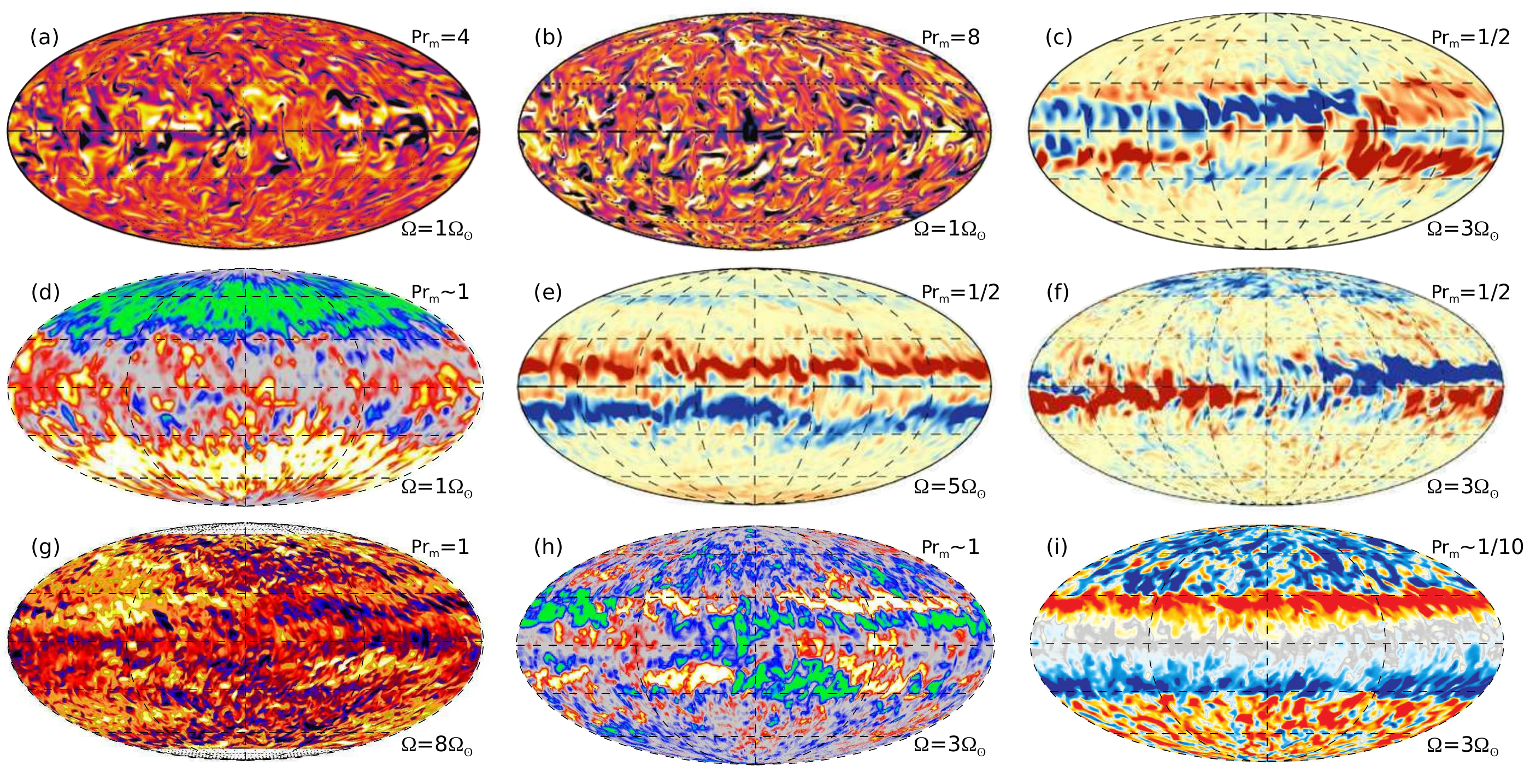}
    \caption{Modern 3-D global-scale convective dynamo simulations, showing snapshots of their
      toroidal magnetic fields in Mollweide projection. The effective magnetic Prandtl number
      ($Pm$) and rotation rate $\Omega$ of each simulation is indicated. (a) First dynamo
      simulation using ASH \citep{brun04}. (b) ASH simulation with a tachocline
      \citep{browning06}. (c) Large-scale, persistent toroidal structures dubbed wreaths maintained
      amidst moderately turbulent convection at lower $Pm$ \citep{brown10}. (d) EULAG solution
      with regular decadal-time-scale magnetic reversals \citep{racine11}. (e) ASH simulation with
      quasi-periodic reversals \citep{brown11a}. (f) Recent continuation of (c) with lower diffusion
      and higher resolution. The greater levels of turbulence permits quasi-periodic cycles to
      emerge \citep{nelson13a}. (g) Cyclic dynamo solution with the Pencil code, showing wreath
      formation in a spherical segment \citep{kapyla13}. (h) EULAG solution rotating at
      $3$~$\Omega_{\odot}$, for comparison with (c), (f), and (i) at the same rotation rate
      \citep{charbonneau13}. (i) Recent simulation using slope-limited diffusion in ASH exhibiting
      very regular cycles. Typical toroidal magnetic field strengths are between 1
      and $10\,$kG in these simulations, with sense of fields suggested by
      color. \label{fig:DynamoMenagerie}}\vspace{-0.25truein}
  \end{figure*} 
\end{center}

%-----------------------------------------------------------------------
\section{Solar-like Stars and Their Cycles} \label{sec:solar}

From the perspective of dynamo theory, lower mass stars fall into the category of large-scale low-$Pm$ dynamos.  Nearly
all such stars are observed to rotate to some degree and to have surface magnetic activity, as discussed above.
Moreover, for sufficiently massive stars, they are likely to support a tachocline that separates their convective
regions from the stable radiative zone below. Simulations of such stars have begun to show how the tachocline can
influence the dynamo in ways that go beyond the original idea of the interface-type dynamo such as those of
\citep{leighton69} and \citep{parker93}.

Having examined some of the basic principles of how differential rotation is built and maintained
and examined how accurately global-scale models can capture such hydrodynamics, the topic of
global-scale dynamo simulations in a solar setting must be addressed. To date, several different
numerical models of solar convective dynamos have achieved cyclical polarity reversals, whether or
not a tachocline is present. In all cases of such cycling global-scale dynamos, the interaction of
convection, rotation, and magnetism are crucial ingredients. Though, precisely how the mechanisms
are arranged so that a solar-like differential rotation can be maintained, and simultaneously yield
a large-scale, periodic dynamo, has yet to be ascertained. Indeed, how the multiply-scaled
convection of a thermally driven system, like the Sun, sempiternally maintains a global-scale kinetic
helicity and sustains the massively disparate scales of the Sun's magnetism is seemingly a mystery.

The earliest attempts to address these issues with global-scale solar dynamo simulations were those of Gilman and
Glatzmaier, who used an anelastic and spherical harmonic based spectral code to simulate the dynamics
\citep{glatzmaier84} rather than the finite difference Boussinesq scheme of \citep{gilman81}. A series of simulations
were laid out in \citep{gilman83} that aimed at assessing how varying the magnetic Prandtl number $Pm$ and the rotation
rate modified the character of the resulting dynamo. In general, it was found that magnetic cycles are a common feature,
provided that there is sufficient energy in the bulk toroidal fields and that the Lorentz force feedback on the
established differential rotation was not too large. Moreover, Gilman also noted that there is little preference for
symmetry or anti-symmetry about the equator and that the toroidal bands of magnetic field tended to propagate toward the
poles. It was also found that, for a fixed Rayleigh number (ratio of thermal driving to diffusion), Taylor number (ratio
of the Coriolis force to viscous diffusion), and thermal Prandtl number (ratio of viscous to thermal diffusion), there
is a peak dynamo efficiency with $Pm\sim 1$. For $Pm$ below unity, the efficiency reached a plateau and for those cases
with $Pm$ above unity it decreased.

\begin{figure*}[t!]
  \begin{center}
    \includegraphics[width=0.9\textwidth]{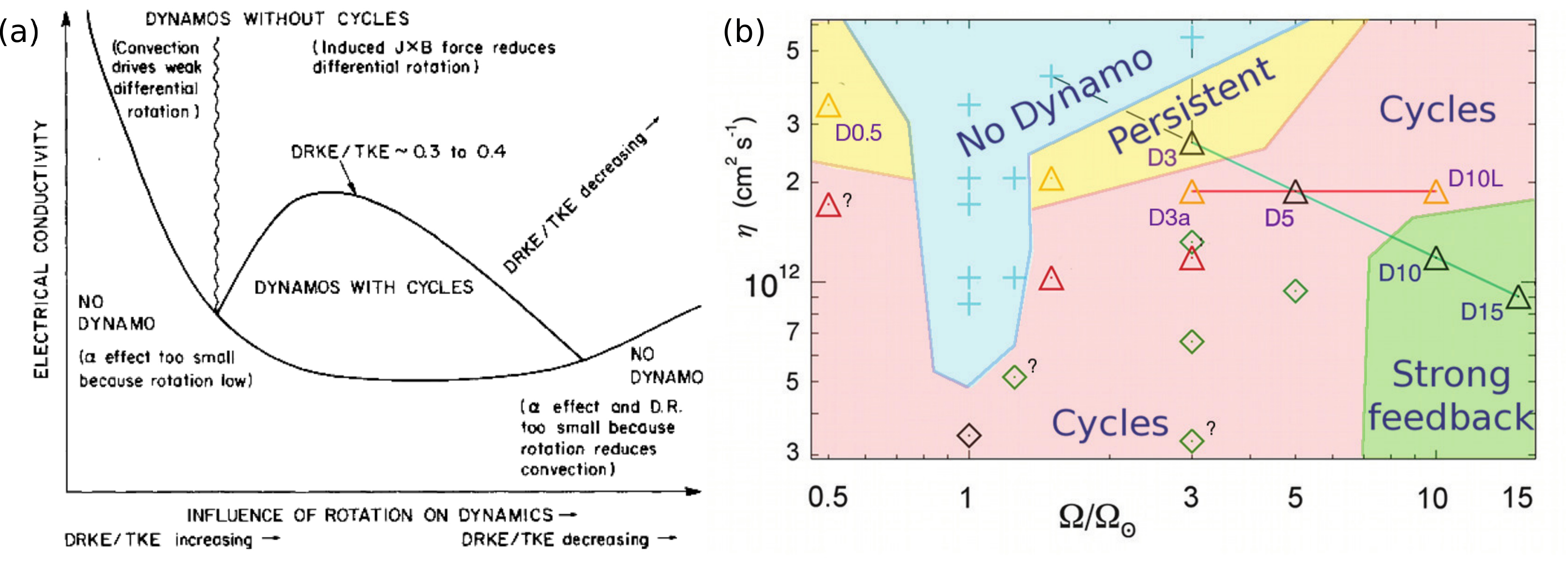}
    \caption[Dynamo regimes old and new]{(a) A sketch of possible dynamo regimes with rotation rate
      and electrical conductivity, illustrating the cycling and persistent dynamo regimes as well as
      those without sustained dynamo action [adapted from \citep{gilman83}]. (b) A set of numerical
      dynamo experiments carried out with varying Prandtl numbers and rotation rates, exhibiting
      many of the behaviors predicted in \citep{gilman83} and some that were not, particularly at
      low rotation rate [adapted from \citep{brown11b}]. \label{fig:dynreg}}\vspace{-0.25truein}
    \end{center}
\end{figure*}

Yet the mismatch between those early simulations and helioseismic inferences effectively ended that line of
investigation for about 15 years. A similar approach to simulating global-scale dynamos was, however, restarted roughly
around 2004 with the inclusion of magnetic fields into the anelastic spherical harmonic (ASH) code (see \citep{brun04}
for instance). These 21st century global-scale convective solar dynamo simulations were carried out at moderate $Pm$,
ranging between 2 and 4. Hence it was much more difficult for a large-scale dynamo to take hold. In keeping with that
notion, the fluctuating fields dominated the magnetic energy, accounting for most of the back-reaction on the flow via
Lorentz forces, whereas the mean fields were relatively weak. There were also no hints of periodic reversals or dynamo
wave behavior. Again as expected with a large $Pm$ dynamo, the magnetic fields exhibit a complex and small-scale spatial
structure and highly-variable time evolution. The radial magnetic field is swept into the downflow lanes, being
stretched and folded there before reaching the resistive scale and reconnecting. The toroidal fields are organized into
somewhat larger scales near the equator appearing as twisted ribbons, whereas they form smaller scales at higher
latitudes (Figure \ref{fig:DynamoMenagerie}a). In \citep{brun04}, it was also found that the convection maintained a solar-like
angular velocity profile despite the influence of Maxwell stresses, which tend to extract energy from the differential
rotation and thus reduce its contrast throughout the convection zone.

The next logical step was to include a stable region and to emulate a tachocline in order to assess its impact upon the
global-scale dynamo \citep{browning06}. Both this simulation and that of \citep{brun04} were fairly laminar with respect
to the scales present in the velocity field. The dynamo action achieved in the convection zone of this simulation is
also quite similar to the earlier calculation of \citep{brun04}, though with an even larger $Pm$. Thus, as one might
expect the scales of magnetism become much smaller than the scales of the convection (Figure
\ref{fig:DynamoMenagerie}b), with the bulk of the magnetic energy appearing in the fluctuating magnetic fields and very
little in the mean fields within the convection zone. The major difference between the dynamo action present in this
simulation, as compared to that \citep{brun04}, is that the stable region provided a reservoir for storing magnetic
energy and the tachocline may have provided a means of generation of large-scale magnetic field. Indeed, within the
stable region, large-scale fields were able to persist, with the majority of the energy being in the mean magnetic
fields. However, there was as yet no large-scale structure that might be able to appear at the surface (i.e.~no
magnetically buoyant structures) nor were there any cyclical polarity reversals.

Rapid progress over the last four years have confirmed the role that convection and rotation likely
play in stellar dynamo action through global-scale dynamo simulations as well as suggested
additional mechanisms that may be present in the solar dynamo. Several research teams have
concurrently made inroads into solar dynamo theory utilizing three distinct numerical techniques,
but nonetheless realizing similar dynamo action. The Anelastic Spherical Harmonic (ASH) code has
provided a framework for constructing and understanding the dynamos that produce and sustain
coherent toroidal fields exhibiting quasi-cyclic behavior. The Eulerian-Lagrangian (EULAG) code has
shown that solutions with very regular cycles can be achieved within its implicit large eddy
simulation (ILES) framework.  Third, the Pencil code that simulates compressive convection has also
found cyclical solutions with large-scale toroidal field structures within its spherical wedge
geometry.

\begin{figure*}[t!]
  \begin{center}
    \includegraphics[width=0.9\textwidth]{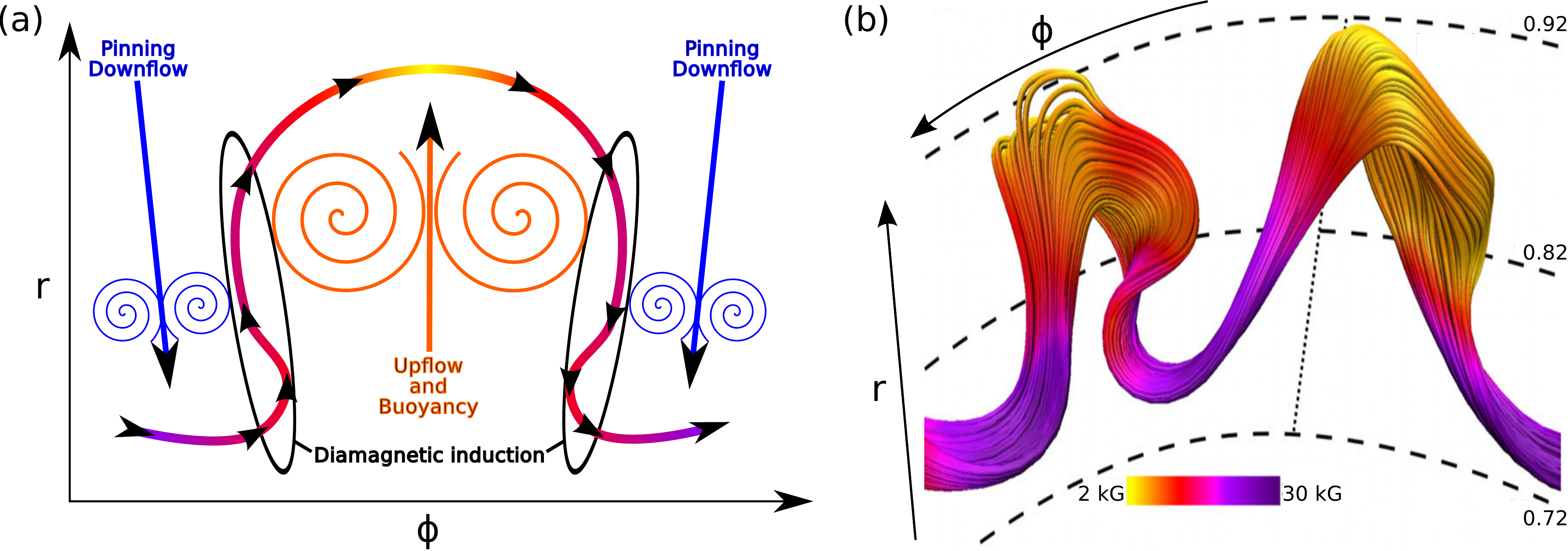}
    \caption[A sketch of buoyant magnetic structures]{(a) A sketch of the processes leading to
      magnetic loop formation in ASH simulations with dynamic Smagorinski: downflows pin the foot
      points of the loops, shear between the upflow and downflow builds a large magnetic diffusivity
      gradient at the edge of the loop leading to diamagnetic induction that builds radial field of
      the correct sign, while advection and buoyancy lift the inner section of the toroidal
      field. (b) Two magnetic loops achieved in an ASH simulation utilizing a dynamic Smagorinski
      scheme shown in a field line rendering [adapted from \citep{nelson11}]. \label{fig:loops}}\vspace{-0.25truein}
    \end{center}
\end{figure*}

\begin{figure*}[t!]
  \begin{center}
    \includegraphics[width=\textwidth]{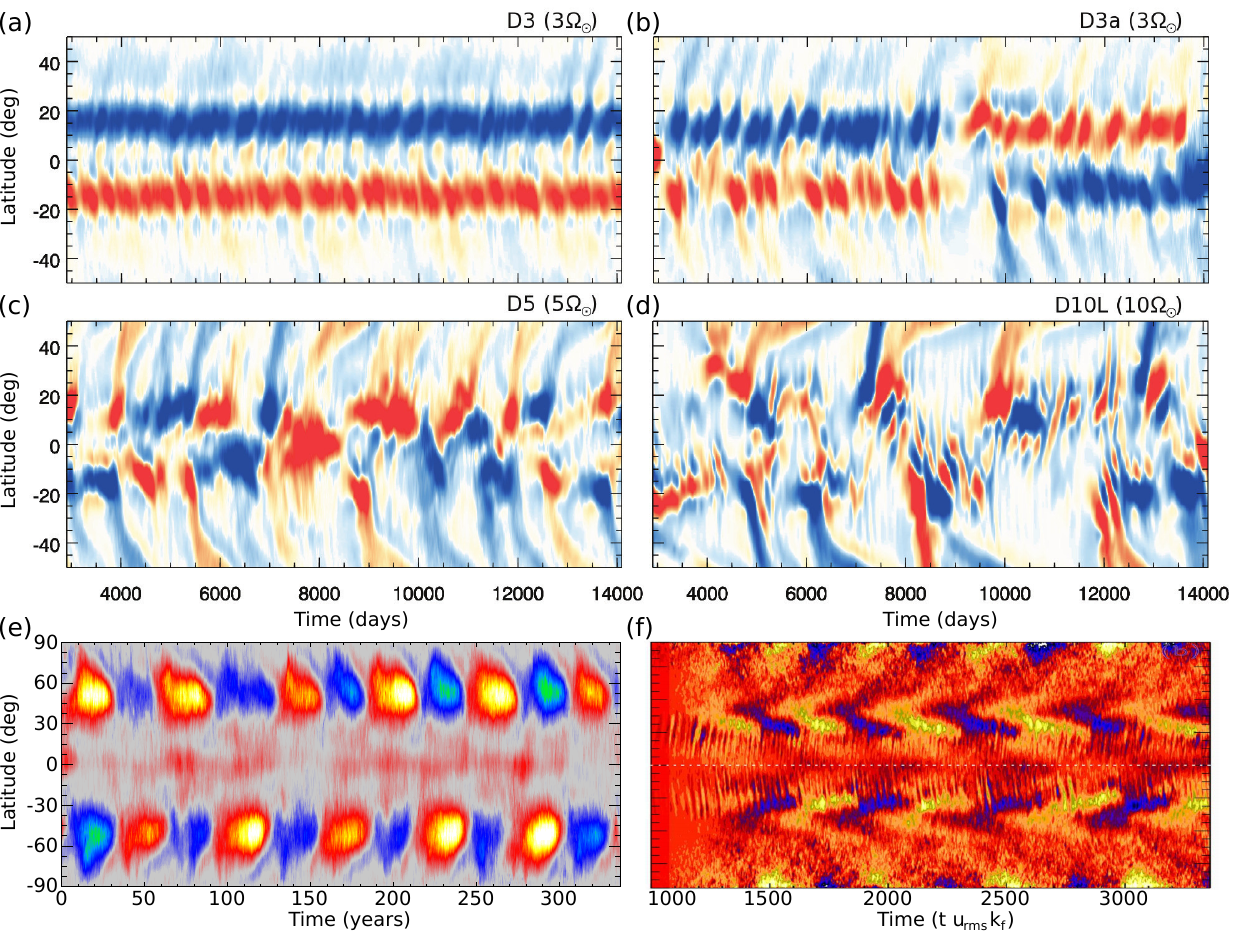}
    \caption[A composite of time-latitude diagrams of toroidal field]{Time and latitude evolution of
      toroidal magnetic fields achieved in several global-scale convective dynamo solutions. (a-d)
      Dynamo calculations carried out with ASH showing four dynamo regimes arising from varying the
      rotation rate and magnetic diffusivity [adapted from \citep{brown11b}]. (e) Toroidal field
      evolution in a solar dynamo simulations utilizing EULAG, showing regular polarity reversals as
      in \citep{racine11}. (f) A cycling case exhibiting polarward and equatorward branches of
      propagating dynamo waves in a Pencil code simulation described in
      \citep{kapyla13}. \label{fig:dyncomp}}\vspace{-0.25truein}
    \end{center}
\end{figure*}

The first 3D nonlinear convective dynamo simulation, beyond those of Gilman and Glatzmaier in the
1980s, that indicated that sustained large-scale toroidal magnetic field structures could be built
within the solar convection zone was presented in \citep{brown10}. A snapshot of this solution's
toroidal magnetic field is shown in Figure \ref{fig:DynamoMenagerie}c and its evolution is shown in
Figure \ref{fig:dyncomp}a. The persistence of such structures had previously been thought
impossible, as one might expect that the highly turbulent convection should rapidly eviscerate it,
transporting the magnetic flux to the boundaries of the domain. What separated this simulation from
those undertaken a few years earlier was its higher rotation rate (or lower Rossby number) and its
lower $Pm$, being about 16 times smaller. The salient point that this simulation brings forth was
that the rate of generation of magnetic field through the mean shear of the differential rotation
could overpower the rate of its destruction by convection and dissipation. Indeed, the convection
instead acts to sustain the differential rotation through Reynolds stresses, which indirectly aids
in the generation of the toroidal fields. The convection then in turn converts the toroidal field
into poloidal field, creating a positive feedback loop that sustains the wreaths against resistive
decay. Although this simulation hinted that perhaps fields could be sustained within the solar
convection zone, it still did not have cyclical behavior. Furthermore, it was still quite laminar,
so the ability to sustain such large-scale magnetic field structures at greater levels of turbulence
and increased complexity of the velocity field was still in question.

That question was partly answered in concurrent dynamo calculations utilizing the EULAG code.  The
ILES implementation of this code allows it to minimize the impacts of diffusion and achieve more
complex velocity and magnetic fields at a given resolution than ASH can achieve. The solar dynamo
modeled with this computational framework exhibited many multi-decadal cycles of roughly 40~years
that were quite regular and persisted for roughly 500~years of evolution despite the fairly complex
flow fields permitted by the low numerical diffusion \citep{ghizaru10, racine11}. A snapshot of the
toroidal magnetic field from this simulation is shown Figure \ref{fig:DynamoMenagerie}d and its
character in time and latitude is illustrated in Figure \ref{fig:dyncomp}e. The turbulent EMF in
these simulations is more amenable to a mean-field theory description than the earlier ASH
simulations, which may in part be due to the greater complexity of the velocity field. Part of the
reason for the greater agreement between these dynamos and mean-field theory may lie in the simple
argument that as the number of convective cells becomes increasingly large the average poloidal
magnetic field generation may approach a mean value. Such a characteristic might be expected if the
averaging operator obeys the central limit theorem, or if the arguments given in \citep{parker55}
hold.

Further exploration of the rotation rate and $Pm$ parameter space using the ASH code continued
apace, with a quasi-periodic cycling dynamo solution published in \citep{brown11a} and a multitude
of solutions explored in \citep{brown11b}. The toroidal fields of the cycling solution are shown in
Figure \ref{fig:DynamoMenagerie}, exhibiting the high degree of longitudinal connectivity that can
be achieved within a low-$Pm$ global-scale dynamo and also lending credence to their moniker of
wreaths. These fields form at or near the regions of the greatest latitudinal shear in the
differential rotation, which is why they tend to be generated closer to the equator. Such
characteristics can be seen in several of the dynamo solutions from \citep{brown11b} as shown in
Figures \ref{fig:dyncomp}a--d. The dynamo behaviors exhibited within the menagerie of simulations in
this latter paper seem to roughly correspond to the regimes described in \citep{gilman83}, as is
illustrated in Figure \ref{fig:dynreg}. The Gilman picture suggests that, as the magnetic
diffusivity is reduced, dynamo action becomes increasingly likely and that cycles become more likely
over a small region. Eventually, however, Lorentz-force feedback quenches both the convection and
differential rotation. Similarly, as the rotation rate is increased, cycles become more likely to a
point and then past this threshold they become less likely due to the rotational suppression of
convection. Yet another set of simulations that sought to determine the influence of the density
scale height and convective driving has recently been carried out that finds distinct transitions
between dipolarity, multi-polarity, and cyclic dynamos \citep{gastine12}, though in less of a solar
context. Indeed, they found that their anelastic dynamo models produced a broad range of magnetic
field geometries, with dipolar configurations for the least stratified and most rapidly rotating and
a gradual transition to multipolar configurations if either the stratification or the rotation rate
was increased. As the density stratification is increased the convective columns in their
simulations become increasingly concentrated in a thin boundary layer near the upper boundary and at
low latitudes. Such patterns of convection tended to be non-axisymmetric solutions. Furthermore, in
multipolar solutions, zonal flows became more significant and led to the production of toroidal
field through the $\Omega$-effect. In those cases, dynamo waves seemed to then play an important
role, leading to polarity reversing dynamos.

Continuing the rapid pace in solar dynamo modeling and under the assumptions of dynamic Smagorinski,
mildly buoyant magnetic structures have been realized in a cycling dynamo simulation \citep[Figure
  \ref{fig:loops};][]{nelson11,nelson13a}. In this fortuitous simulation, the dynamic Smagorinski
diffusion permits a greater degree of turbulence than can be achieved in the eddy viscosity
model. It helps to establish a pattern of quasi-buoyant magnetic structures that form within the
convection zone, which is in contrast to the common wisdom that they could only form in the
tachocline as suggested in \citep{parker93}. Regardless of how these structures come about, they
tend to have statistics strikingly similar to the statistics of solar observations regarding Joy's
and Hale's laws \citep{nelson13b}. These magnetic structures arise from the structure of the flow
field, natural buoyancy, and partly from diamagnetic processes.  The latter process has not yet been
fully considered, so the logic of it will be expanded upon a bit here. The downflows help to pin
down the foot points of the loop-like magnetic fields, whereas the upflow between two downflows
advects a portion of the magnetic field toward the upper boundary of the simulation. The natural
magnetic and thermal buoyancy of the structure further boost the forces, compelling it to
rise.

There have been recent simulations that also exhibit two branches of propagating nonlinear dynamo waves one that is
equatorward at low latitudes and another that is poleward at higher latitudes \citep[Figure
\ref{fig:dyncomp}f;][]{kapyla13}. As was also seen in \citep{gastine12}, they find that the density stratification of
the simulation has a substantial impact on the overall character of the dynamo. In particular, one of their solutions
exhibits both equatorward and poleward propagating dynamo waves (Figure \ref{fig:dyncomp}f), though the reason for this
is unclear. The propagation characteristics of the dynamo are attributed to the stratification. Yet the number of
density scale heights that are simulated in these calculations is only about 3 to 4.5 density scale heights, which is
similar to those of previous ASH simulations. It may be that the differential rotation has a region of near-surface
shear, which can promote equatorward propagation provided that the sign of the kinetic helicity does not reverse at the
same time.  One study has shed further light on this subject by assessing the role of the columnarity of the convective
structures in determining the direction of propagation of a dynamo wave \citep{duarte16}.  In that paper, below a
threshold columnarity of about 0.3, an equatorward propagating dynamo wave is established. More precisely, in some of
their simulations, the direction of propagation of dynamo waves is altered primarily by an inversion of the kinetic
helicity in the deep interior relative to the near-surface kinetic helicity, rather than by changes in the differential
rotation. This inversion tends to occur in cases with a low Prandtl number, internal heating, and in regions where the
local density gradient is relatively small.

Several recent convective dynamo simulations carried out in spherical segments with the Pencil code have also manifested
unprecedented magnetic self-organization, with strong mean field generation in nonlinear dynamo regimes, and cyclic
activity on decadal time scales \citep[Figure \ref{fig:DynamoMenagerie}g;][]{kapyla12}.  One simulation in particular, a
global-scale simulation using the ASH code and slope-limited diffusion, achieved a convective dynamo that exhibits many
of these interesting dynamo features, generating magnetic fields with a prominent polarity cycle occurring roughly every
6.2 years for several hundred years of evolution \citep[Figure \ref{fig:DynamoMenagerie}i;][]{augustson15,miesch16}. The
polarity reversals seen in that simulation are related to the variation of the differential rotation and to a resistive
collapse of the large-scale magnetic field. An equatorial migration of the magnetic field is also seen, which in turn is
due to the strong, nonlinear modulation of the differential rotation rather than a dynamo wave. This simulation also
enters an interval with reduced magnetic energy at low latitudes lasting roughly 16 years (about 2.5 polarity cycles),
during which the polarity cycles are disrupted and after which the dynamo recovers its regular polarity cycles. An
analysis of this grand minimum reveals that it likely arises through the interplay of symmetric and antisymmetric dynamo
families. This intermittent dynamo state potentially results from the simulation’s relatively low magnetic Prandtl
number.

Another set of Pencil code simulations have begun to investigate how the upper boundary can influence the operation
of the dynamo, such as through the release of magnetic helicity \citep[e.g.,][]{brandenburg05}. The generation of
large-scale fields in dynamos is often associated with an upscale transfer of magnetic helicity. In order to sustain
this transfer, small-scale helicity must be either dissipated or removed from the system by passing through the
boundaries. A possible manifestation of this on the Sun is that the magnetic helicity released in coronal mass ejections
\citep{blackman03}. Indeed, these recent simulations have demonstrated that convective dynamos can eject helicity
through CME-like eruptions and separately form bipolar magnetic features akin to sunspots
\citep{warnecke12,warnecke13a,warnecke13b}, yet it is still unclear how such events effect the operation of the dynamo.

Finally, \citep{guerrero16} presented a suite of global-scale simulations of rotating turbulent convective dynamos that
focused on determining the role of the tachocline in setting the cycle period. The first set of their simulations
considered only a stellar convective envelope, whereas the second set aimed at establishing a tachocline, and so they
also included the upper part of a radiative zone. For the first set of models, either oscillatory or steady dynamo
solutions are obtained, depending upon the global convective Rossby number of the simulation in question. The models in
the second set naturally develop a tachocline, leading to the generation of a strong mean magnetic
field. Since the field is also deposited in the stable deeper layer, its evolutionary timescale is much longer than in
the models without a tachocline. Due to non-local dynamo effects, the magnetic field in the upper turbulent convection
zone evolves on the same timescale as the deep field. These models result in either an oscillatory dynamo with a 30 yr
period or a steady dynamo depending again on Rossby number.  Hence, the presence of a tachocline appears to lengthen the
cycle period.

\section{Intermediate-mass Stars and Convective Core Dynamos} \label{sec:massive}

There may be the potential to identify a few regimes for which some global-scale aspects of stellar dynamos might be
estimated with only a knowledge of the basic parameters of the system. For example, consider how the magnetic energy
contained in the system may change with a modified level of turbulence (or stronger driving), or how does the ratio of
the dissipative length scales impact that energy, or how does rotation influence it? Some of these questions will be
addressed here. Establishing the global-parameter scalings of convective dynamos, particularly with stellar mass and
rotation rate, is useful given that they provide an order of magnitude approximation of the magnetic field strengths
generated within the convection zones of stars as they evolve from the pre-main-sequence to a terminal phase. This in
turn permits the placement of better constraints upon transport processes, such as those for elements and angular
momentum, most of which occur over structurally-significant evolutionary timescales.  This could be especially useful in
light of the recent evidence for magnetic fields within the cores of red giants, pointing to the existence
of a strong core dynamo being active in a large fraction of intermediate mass stars \citep{fuller15,cantiello16}.

\section{Scaling of Magnetic and Kinetic Energies} \label{sec:forcescaling}

Convective flows often possess distributions of length scales and speeds that are peaked near a
single characteristic value. One simple method to estimate these quantities is to divine that the
energy containing flows have roughly the same length scale as the depth of the convection zone and
that the speed of the flows is directly related to the rate of energy injection (given here by the
stellar luminosity) and inversely proportional to the density of the medium into which that energy
is being injected \citep{augustson12}. The latter is encapsulated as $v_{rms} \propto
\left(2L/\rho_{\mathrm{CZ}}\right)^{1/3}$, where $\rho_{\mathrm{CZ}}$ is the average density in the
convection zone. However, such a mixing-length velocity prescription only provides an order of
magnitude estimate as the precise level of equipartition depends sensitively upon the dynamics
\citep[e.g.,][]{yadav16}. Since stars are often rotating fairly rapidly, taking for instance young
low-mass stars and most intermediate and high-mass stars, their dynamos may reach a
quasi-magnetostrophic state wherein the Coriolis acceleration also plays a significant part in
balancing the Lorentz force.  Such a balance has been addressed and discussed at length in
\citep{christensen10, brun15}, and \citep{augustson17} for instance.

To characterize the force balance, consider the compressive MHD Navier-Stokes vorticity equation, wherein one has that

\vspace{-0.25truein}
\begin{center}
  \begin{align}
    \frac{\partial\boldsymbol{\omega}}{\partial t} &= \curl{\left[\vv\cross\boldsymbol{\omega_P}+\frac{1}{\rho}\left(\frac{\vJ\cross\vB}{c}+\dvg{\sigma}-\grad P\right)\right]},
  \end{align}
\end{center}

\noindent where $\vv$ is the velocity, $\vB$ is the magnetic field, $\rho$ is the density, $P$ the pressure. The
following variables are also defined, with the current being $\vJ = c\curl{\vB}/4\pi$, with $c$ being the speed of
light, $\sigma$ being the viscous stress tensor, and where $\boldsymbol{\omega} = \curl{\vv}$ and
$\boldsymbol{\omega_P}=2\bomega+\boldsymbol{\omega}$.

Taking the dot product of this equation with $\boldsymbol{\omega}$ gives rise to the equation for
the evolution of the enstrophy.  Integrating that equation over the volume of the convective domain
and over a reasonable number of dynamical times such that the system is statistically steady yields

\vspace{-0.25truein} 
\begin{center}
  \begin{align}
    &\!\!\int\!\! d\mathbf{S}\!\bcdot\!\left[\vv\cross\boldsymbol{\omega_P}+\frac{1}{\rho}\left(\frac{\vJ\cross\vB}{c}+\dvg{\sigma}-\grad P\right)\right]\cross\boldsymbol{\omega} \nonumber \\
    &\!\!+\!\!\int\!\! dV \left(\curl{\boldsymbol{\omega}}\right)\!\bcdot\!\left[\vv\cross\boldsymbol{\omega_P}+\frac{1}{\rho}\left(\frac{\vJ\cross\vB}{c}+\dvg{\sigma}-\grad P\right)\right]\!=\!0.
  \end{align}
\end{center}

\noindent If no enstrophy is lost through the boundaries of the convective domains, then the surface
integral vanishes, leaving

\vspace{-0.25truein} 
\begin{center}
  \begin{align}
    \!\!\int\!\! dV \left(\curl{\boldsymbol{\omega}}\right)\!\bcdot\!\left[\vv\cross\boldsymbol{\omega_P}+\frac{1}{\rho}\left(\frac{\vJ\cross\vB}{c}+\dvg{\sigma}-\grad P\right)\right]=0.
  \end{align}
\end{center}

\noindent This assumption effectively means that magnetic stellar winds will not be part of this
scaling analysis.  For timescales consistent with the dynamical timescales of the dynamos considered
here, this is a reasonably valid assumption. Then, since $\boldsymbol{\nabla\!\times\!\omega}$ is
not everywhere zero, the terms in square brackets must be zero, which implies that

\vspace{-0.25truein} 
\begin{center}
  \begin{align}
    \rho\vv\cross\boldsymbol{\omega_P}+\frac{\vJ\cross\vB}{c}+\dvg{\sigma}-\grad P=0.
  \end{align}
\end{center}

\noindent Taking the curl of this equation eliminates the pressure contribution and gives

\vspace{-0.25truein}
\begin{center}
  \begin{align}
    \curl{\left[\rho\vv\cross\boldsymbol{\omega}+2\rho\vv\cross\bomega+\frac{\vJ\cross\vB}{c}+\dvg{\sigma}\right]}=0.\label{eqn:curlforce}
  \end{align}
\end{center}

\noindent This, then, is the primary balance between inertial, Coriolis, Lorentz, and viscous
forces.  If one again scales the derivatives as the inverse of a characteristic length scale $\ell$,
and takes fiducial values for the other parameters, the scaling relationship for the above equation
gives

\vspace{-0.25truein} 
\begin{center}
  \begin{align}
    \rho \mathrm{v_{rms}^2}/\ell^2 + 2\rho \mathrm{v_{rms}}\Omega_0/\ell+ B^2/4\pi\ell^2 + \rho\nu \mathrm{v_{rms}}/\ell^3 \approx 0,
  \end{align}
\end{center}

\noindent which when divided through by $\rho\mathrm{v_{rms}^2}/\ell^2$ yields

\vspace{-0.25truein} 
\begin{center}
  \begin{align}
    \mathrm{ME/KE} \propto 1 + Re^{-1} + Ro^{-1}.
  \end{align}
\end{center}

\noindent Here the Reynolds number is taken to be $Re = \mathrm{v_{rms}}\ell/\nu$.  However, the leading term of this
scaling relationship is found to be less than unity, at least when assessed through simulations.  Hence, it should be
replaced with a parameter to account for dynamos that are subequipartition. This leaves the following

\vspace{-0.25truein} 
\begin{center}
  \begin{align}
    \mathrm{ME/KE} \propto \beta(Ro,Re) + Ro^{-1}, \label{eqn:forcescaling}
  \end{align}
\end{center}

\noindent where $\beta(Ro,Re)$ is unknown apriori as it depends upon the intrinsic ability of the
non-rotating system to generate magnetic fields, which in turn depends upon the specific details of
the system such as the boundary conditions and geometry of the convection zone.

Thus, for a subset of dynamos such as those discussed in \citep{augustson16}, Equation
(\ref{eqn:forcescaling}) may hold and the ratio of the total magnetic energy (ME) to the kinetic
energy (KE) will depend upon the inverse Rossby number plus a constant offset, which will be
sensitive to details of the dynamics. In some circumstances, the constant offset may also depend
upon the Rossby number. Hence, such dynamos are sensitive to the degree of rotational constraint on
the convection and upon the intrinsic ability of the convection to generate a sustained dynamo.
Moreover, this line of argument indicates that the inertial term may give rise to a minimum magnetic
energy state. This allows a bridge between two dynamo regimes, the equipartition slowly rotating
dynamos and the rapidly rotating magnetostrophic regime, where $\mathrm{ME/KE}\propto Ro^{-1}$. For
low Rossby numbers, or large rotation rates, it is possible that the dynamo can reach
superequipartition states where $\mathrm{ME/KE}>1$. Indeed, it may be much greater than unity, as is
expected for the Earth's dynamo (see Figure 6 of \citep{roberts13}).

\section{Comparison of Scaling Relationships}

\begin{figure}[!t]
  \centering
  \includegraphics[width=0.95\linewidth]{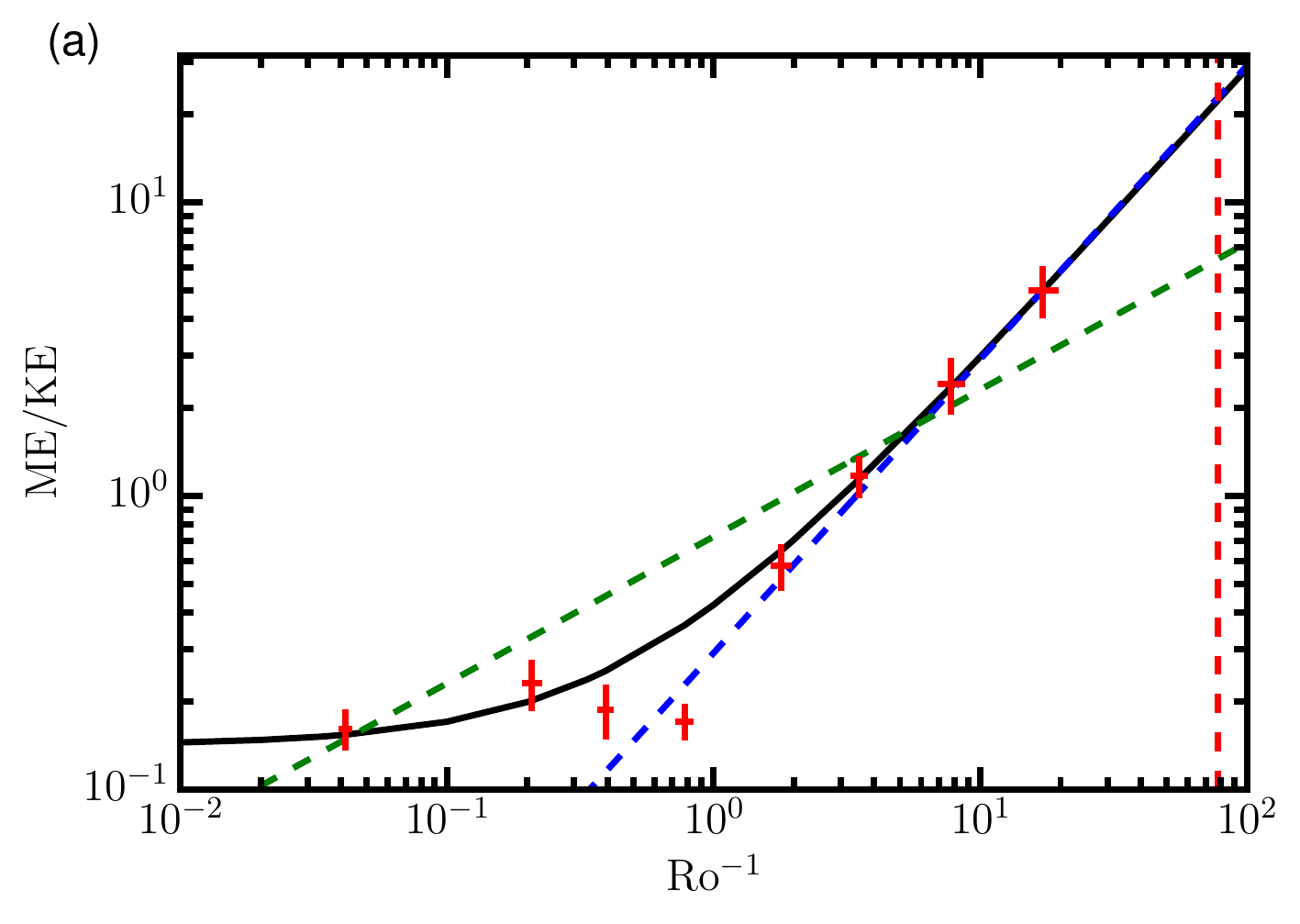}
  \caption{The scaling of the ratio of magnetic to kinetic energy ($\mathrm{ME/KE}$) with inverse Rossby number
    ($Ro^{-1}$). The black curve indicates the scaling defined in Equation (\ref{eqn:forcescaling}), with
    $\beta=0.5$. The blue dashed line is for magnetostrophy, e.g. $\beta=0$.  The green dashed line is that for
    buoyancy-work-limited dynamo scaling, which requires $\mathrm{ME/KE}\propto Ro^{-1/2}$. The red dashed line
    indicates the critical Rossby number of the star, correspoding to its rotational breakup velocity. The uncertainty
    of the measured Rossby number and energy ratio that arises from temporal variations are indicated by the size of the
    cross for each data point.}
  \label{fig:scaling}\vspace{-0.25truein}
\end{figure}

\begin{figure}[ht]
  \centering
  \includegraphics[width=0.95\linewidth]{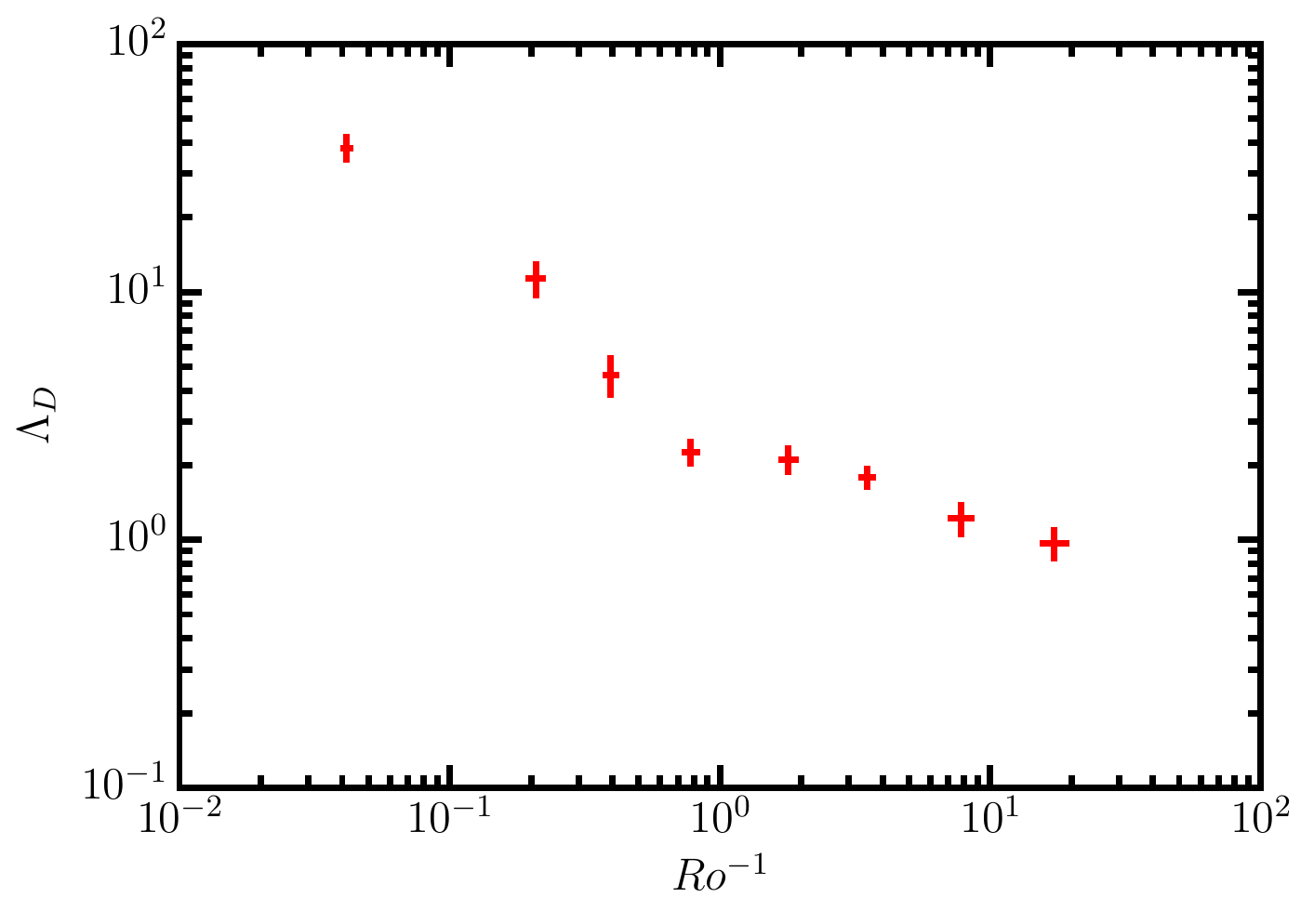}
  \caption{The scaling of the dynamic Elsasser number ($\Lambda_D$) with inverse Rossby number
    ($Ro^{-1}$). The uncertainty of the measured Rossby number and dynamic Elsasser number that
    arises from temporal variations are indicated by the size of the cross for each data point.}\vspace{-0.25truein}
  \label{fig:elsasser}
\end{figure}

Since we wish to compare these three schemes, consider the data for the evolution of a set of MHD
simulations using the Anelastic Spherical Harmonic code presented in \citep{augustson16}. These
simulations attempt to capture the dynamics within the convective core of a massive star. In such
stars, $Pm$ is approximately four throughout the core, falls to about two at the core boundary, and
to $1/10$ nearer to the stellar surface in the radiative exterior.  Hence, the convective core can
be considered as a large magnetic Prandtl number dynamo. Such $Pm$ regimes are accessible from a
numerical point of view.  These convective cores also present an astrophysical dynamo in which
large-eddy simulations can more easily capture the hierarchy of relevant diffusive timescales. This
is especially so given that the density contrast across the core is generally small, being a bit
less than two.  The heat generation from nuclear burning processes deep within the core and the
radiative cooling nearer the radiative zone are also quite smoothly distributed in radius. So, there
are no inherent difficulties with resolving internal boundary layers.

In the suite of simulations presented in \citep{augustson16}, the rotation rates employed lead to
nearly three decades of coverage in Rossby number, as shown in Figure (\ref{fig:scaling}). In that
figure, the force-based scaling derived in \S\ref{sec:forcescaling} depicted by the black curve
(where $\mathrm{ME/KE}\propto 0.5+Ro^{-1}$) does a reasonably good job of describing the nature of
the superequipartition state for a given Rossby number. Note, however, that the constant of
proportionality in Equation (\ref{eqn:forcescaling}) has been determined using the data itself,
which is true also of the other two scaling laws shown in Figure (\ref{fig:scaling}).

These simulated convective core dynamos appear to enter a regime of magnetostrophy for the four
cases with the lowest average Rossby number, where the scaling for the magnetostrophic regime is
denoted by the dashed blue line in Figure (\ref{fig:scaling}).  This transition to the magnetostrophic
regime is further evidenced in Figure (\ref{fig:elsasser}), which shows the dynamic Elsasser number

\vspace{-0.25truein}
\begin{center}
  \begin{align}
    \Lambda_D = \mathrm{B_{rms}}^2/(8\pi\rho_0\Omega_0\mathrm{v_{rms}}\ell),
  \end{align}
\end{center}

\noindent where $\ell$ is the typical length scale of the current density $\vJ$. Indeed, as
$\Lambda_D$ approaches unity, the balance between the Lorentz and the Coriolis forces also
approaches unity, indicating that the dynamo is very close to magnetostrophy.

The scaling law derived in \citep{davidson13} and \citep{augustson17} suggests that there is a direct link between the
buoyancy work and ohmic dissipation, whereas the Coriolis force only requires that there be two integral length scales:
one for flows parallel and another for flows perpendicular to the rotation axis.  Following this logic, one can show
that this requires that the level of superequipartition of the magnetic energy relative to the convective kinetic energy
is proportional to $Ro^{-1/2}$. This scaling does not capture the behavior of the set of dynamos in \citep{augustson16}
very well, as indicated by the lack of agreement between the blue line and the data points in Figure
(\ref{fig:scaling}). In contrast, for the many dynamo simulations and data shown in \citep{christensen09} and
\citep{christensen10}, it does perform well. This latter fact could be related to the fact that for large $Pm$ the
buoyancy work potentially has an additional $Ro$ dependence.

\section{Conclusions} \label{sec:con}

This conference proceeding has attempted to shine some light on the link between many observations and dynamo theory.
In particular, there was a focus on the changing nature of the observed magnetism as a function of mass and rotation
rate, linking Ca II, H-$\alpha$, and X-ray observations and magnetic activity.  On the theory side, some effort was spent
explaining the existence of two kinds of dynamos that depend upon magnetic Prandtl number and that they have relevance
to stellar astrophysics.  In particular, it was shown that there is likely a dichotomy in the kind of dynamo action
taking place within stars that possess a convective core and those that possess an exterior convective envelope. Recent
advances in describing the dynamos of solar-like stars was covered in a fair level of detail. Finally, there is a brief
section discussing scaling laws for the level of the partitioning of magnetic energy and kinetic energy, which is
relevant for intermediate and massive stars that possess a convective core.

In attempting to explain the great many observations of stellar magnetism, including its variability and intensity,
dynamo theory has made some surprising advances in unexpected directions. Numerical simulations in particular have
provided a fruitful path to understanding many of the observed features of the solar dynamo.  While the exact processes
that set the cycle period remain unknown, it has been shown that the tachocline can play a significant role in
recovering longer cycle periods \citep{guerrero16}. Most importantly, however, cyclic solutions seem now to be a more
robust feature than in the past due to increased computational power and thus increased levels of
turbulence. Further, the equatorward and poleward propagation of surface features, provided that they are tied to
the deeper dynamics in a significant way, may follow a deeper dynamo wave, weather it is linear or nonlinear in origin
\citep{kapyla13,augustson15,duarte16}.  Though promising, all of these recent convective dynamo simulations are still in
the early stages of discovery. They are still quite sensitive to the treatment of diffusion where slight changes in
the form and magnitudes of the diffusivities can lead to quite different behavior. Moreover, the physical mechanisms
responsible for determining the cycle properties established within these various convection models remains
unknown. What is clear is that these cyclic dynamos operate very differently than typical mean field dynamo models in
that the meridional flow plays a minimal role in regulating the cycle period and flux emergence plays little or no role
in poloidal field generation.

There appears to be two scaling laws for the level of equipartition of magnetic and kinetic energy that are applicable to
stellar systems, one in the high magnetic Prandtl number regime and another in the low magnetic Prandtl number regime.
Within the context of the large magnetic Prandtl number systems, the magnetic energy of the system scales as the kinetic
energy multiplied by an expression that depends upon an offset, which depends upon the details of the non-rotating
system, e.g. on Reynolds number, Rayleigh number, and Prandtl number, plus the inverse of the convective Rossby number,
as was considered in \citep{augustson16}. For low magnetic Prandtl number and fairly rapidly rotating systems, such as
the geodynamo and rapidly rotating low-mass stars, another scaling law that relies upon a balance of buoyancy work and
magnetic dissipation as well as a balance between the buoyancy, Coriolis, and Lorentz forces may be more
applicable. When focused on in detail, this yields a magnetic energy that scales as the kinetic energy multiplied by the
inverse square root of the convective Rossby number, as has been shown to be widely applicable in such low Prandtl
number and low Rossby number systems \citep{davidson13}. On the side of numerical experiments, a larger range of
Reynolds number and level of supercriticality needs to be explored. Indeed, in \citep{yadav16}, some work has already
attempted to do this for the geodynamo.

\subsection*{Acknowledgement}{K.~C. Augustson acknowledges support from the the ERC SPIRE 647383
grant.}

%-----------------------------------------------------------------------

\begin{thebibliography}{90}

\bibitem{howe09}
R.~Howe, Liv. Rev. Sol. Phys. \textbf{6} (2009)

\bibitem{brun15}
A.S. {Brun}, M.K. {Browning}, M.~{Dikpati}, H.~{Hotta}, A.~{Strugarek}, Space
  Sci. Rev. \textbf{196}, 101 (2015)

\bibitem{zahn91}
J.P. {Zahn}, A\&A \textbf{252}, 179 (1991)

\bibitem{gough98}
D.O. {Gough}, M.E. {McIntyre}, Nature \textbf{394}, 755 (1998)

\bibitem{zhao13}
J.~{Zhao}, R.S. {Bogart}, A.G. {Kosovichev}, T.L. {Duvall}, Jr., T.~{Hartlep},
  ApJL \textbf{774}, L29 (2013)

\bibitem{schmitt85}
J.H.M.M. {Schmitt}, L.~{Golub}, F.R. {Harnden}, Jr., C.W. {Maxson},
  R.~{Rosner}, G.S. {Vaiana}, ApJ \textbf{290}, 307 (1985)

\bibitem{berghoefer97}
T.W. {Bergh\"{o}fer}, J.H.M.M. {Schmitt}, R.~{Danner}, J.P. {Cassinelli}, A\&A
  \textbf{322}, 167 (1997)

\bibitem{sana06}
H.~{Sana}, G.~{Rauw}, Y.~{Naz{\'e}}, E.~{Gosset}, J.M. {Vreux}, MNRAS
  \textbf{372}, 661 (2006)

\bibitem{antokhin08}
I.I. {Antokhin}, G.~{Rauw}, J.M. {Vreux}, K.A. {van der Hucht}, J.C. {Brown},
  A\&A \textbf{477}, 593 (2008)

\bibitem{wrightn11}
N.J. {Wright}, J.J. {Drake}, E.E. {Mamajek}, G.W. {Henry}, ApJ \textbf{743}, 48
  (2011)

\bibitem{dalsgaard04}
J.~{Christensen-Dalsgaard}, Sol. Phys. \textbf{220}, 137 (2004)

\bibitem{dupret09}
M.A. {Dupret}, K.~{Belkacem}, R.~{Samadi}, J.~{Montalban}, O.~{Moreira},
  A.~{Miglio}, M.~{Godart}, P.~{Ventura}, H.G. {Ludwig}, A.~{Grigahc{\`e}ne}
  et~al., A\&A \textbf{506}, 57 (2009)

\bibitem{bedding11}
T.R. {Bedding}, B.~{Mosser}, D.~{Huber}, J.~{Montalb{\'a}n}, P.~{Beck},
  J.~{Christensen-Dalsgaard}, Y.P. {Elsworth}, R.A. {Garc{\'{\i}}a},
  A.~{Miglio}, D.~{Stello} et~al., Nature \textbf{471}, 608 (2011)

\bibitem{mosser14}
B.~{Mosser}, O.~{Benomar}, K.~{Belkacem}, M.J. {Goupil}, N.~{Lagarde},
  E.~{Michel}, Y.~{Lebreton}, D.~{Stello}, M.~{Vrard}, C.~{Barban} et~al., A\&A
  \textbf{572}, L5 (2014)

\bibitem{beck12}
P.G. {Beck}, J.~{De Ridder}, C.~{Aerts}, T.~{Kallinger}, S.~{Hekker}, R.A.
  {Garc{\'{\i}}a}, B.~{Mosser}, G.R. {Davies }, Astro. Nachr. \textbf{333}, 967
  (2012)

\bibitem{deheuvels15}
S.~{Deheuvels}, J.~{Ballot}, P.G. {Beck}, B.~{Mosser}, R.~{{\O}stensen}, R.A.
  {Garc{\'{\i}}a}, M.J. {Goupil}, A\&A \textbf{580}, A96 (2015)

\bibitem{fuller15}
J.~{Fuller}, M.~{Cantiello}, D.~{Stello}, R.A. {Garcia}, L.~{Bildsten}, Science
  \textbf{350}, 423 (2015)

\bibitem{cantiello16}
M.~{Cantiello}, J.~{Fuller}, L.~{Bildsten}, ApJ \textbf{824}, 14 (2016)

\bibitem{peres00}
G.~{Peres}, S.~{Orlando}, F.~{Reale}, R.~{Rosner}, H.~{Hudson}, ApJ
  \textbf{528}, 537 (2000)

\bibitem{herschel1795}
W.~{Herschel}, Roy. Soc. London Phil. Trans. Series I \textbf{85}, 46 (1795)

\bibitem{herschel1796}
W.~{Herschel}, Royal Society of London Philosophical Transactions Series I
  \textbf{86}, 452 (1796)

\bibitem{abney1877a}
W.D.W. {Abney}, MNRAS \textbf{37}, 278 (1877)

\bibitem{shajn29}
G.~{Shajn}, O.~{Struve}, MNRAS \textbf{89}, 222 (1929)

\bibitem{wilson66}
O.C. {Wilson}, ApJ \textbf{144}, 695 (1966)

\bibitem{stassun11}
K.G. {Stassun}, L.~{Hebb}, K.~{Covey}, A.A. {West}, J.~{Irwin}, R.~{Jackson},
  M.~{Jardine}, J.~{Morin}, D.~{Mullan}, I.N. {Reid}, 17th Cool Stars Proceedings, \textbf{448}, 505 (2011)

\bibitem{wilson63}
O.C. {Wilson}, ApJ \textbf{138}, 832 (1963)

\bibitem{moffatt78}
H.K. {Moffatt}, \emph{{Magnetic field generation in electrically conducting
  fluids}} (Cambridge, England, Cambridge University Press.~353 p., 1978)

\bibitem{skumanich72}
A.~{Skumanich}, ApJ \textbf{171}, 565 (1972)

\bibitem{vaughan80}
A.H. {Vaughan}, G.W. {Preston}, Pub. Astron. Soc. Pac. \textbf{92}, 385 (1980)

\bibitem{soderblom93a}
D.R. {Soderblom}, J.R. {Stauffer}, J.D. {Hudon}, B.F. {Jones}, ApJS
  \textbf{85}, 315 (1993)

\bibitem{wrightj04}
J.T. {Wright}, G.W. {Marcy}, R.P. {Butler}, S.S. {Vogt}, ApJS \textbf{152}, 261
  (2004)

\bibitem{pallavicini81}
R.~{Pallavicini}, L.~{Golub}, R.~{Rosner}, G.S. {Vaiana}, T.~{Ayres}, J.L.
  {Linsky}, ApJ \textbf{248}, 279 (1981)

\bibitem{vilhu84}
O.~{Vilhu}, A\&A \textbf{133}, 117 (1984)

\bibitem{micela85}
G.~{Micela}, S.~{Sciortino}, S.~{Serio}, G.S. {Vaiana}, J.~{Bookbinder},
  L.~{Golub}, F.R. {Harnden}, Jr., R.~{Rosner}, ApJ \textbf{292}, 172 (1985)

\bibitem{vilhu87}
O.~{Vilhu}, F.M. {Walter}, ApJ \textbf{321}, 958 (1987)

\bibitem{pizzolato03}
N.~{Pizzolato}, A.~{Maggio}, G.~{Micela}, S.~{Sciortino}, P.~{Ventura}, A\&A
  \textbf{397}, 147 (2003)

\bibitem{jardine99}
M.~{Jardine}, Y.C. {Unruh}, A\&A \textbf{346}, 883 (1999)

\bibitem{blackman15}
E.G. {Blackman}, J.H. {Thomas}, MNRAS \textbf{446}, L51 (2015)

\bibitem{blackman16}
E.G. {Blackman}, J.E. {Owen}, MNRAS \textbf{458}, 1548 (2016)

\bibitem{schmidt15}
S.J. {Schmidt}, S.L. {Hawley}, A.A. {West}, J.J. {Bochanski}, J.R.A.
  {Davenport}, J.~{Ge}, D.P. {Schneider}, Astron. J. \textbf{149}, 158 (2015)

\bibitem{spiegel80}
E.A. {Spiegel}, N.O. {Weiss}, Nature \textbf{287}, 616 (1980)

\bibitem{parker93}
E.N. {Parker}, ApJ \textbf{408}, 707 (1993)

\bibitem{west08}
A.A. {West}, S.L. {Hawley}, J.J. {Bochanski}, K.R. {Covey}, I.N. {Reid},
  S.~{Dhital}, E.J. {Hilton}, M.~{Masuda}, Astron. J. \textbf{135}, 785 (2008)

\bibitem{jardine02}
M.~{Jardine}, A.~{Collier Cameron}, J.F. {Donati}, MNRAS \textbf{333}, 339
  (2002)

\bibitem{vidotto14}
A.A. {Vidotto}, M.~{Jardine}, J.~{Morin}, J.F. {Donati}, M.~{Opher}, T.I.
  {Gombosi}, MNRAS \textbf{438}, 1162 (2014)

\bibitem{reville15}
V.~{R{\'e}ville}, A.S. {Brun}, A.~{Strugarek}, S.P. {Matt}, J.~{Bouvier}, C.P.
  {Folsom}, P.~{Petit}, ApJ \textbf{814}, 99 (2015)

\bibitem{braginskii65}
S.I. {Braginskii}, Reviews of Plasma Physics \textbf{1}, 205 (1965)

\bibitem{schekochihin04}
A.A. {Schekochihin}, S.C. {Cowley}, S.F. {Taylor}, J.L. {Maron}, J.C.
  {McWilliams}, ApJ \textbf{612}, 276 (2004)

\bibitem{schekochihin07}
A.A. {Schekochihin}, A.B. {Iskakov}, S.C. {Cowley}, J.C. {McWilliams}, M.R.E.
  {Proctor}, T.A. {Yousef}, New J. of Physics \textbf{9}, 300 (2007)

\bibitem{brandenburg09}
A.~{Brandenburg}, ApJ \textbf{697}, 1206 (2009)

\bibitem{parker55}
E.N. {Parker}, ApJ \textbf{122}, 293 (1955)

\bibitem{backus58}
G.~{Backus}, Annals of Physics \textbf{4}, 372 (1958)

\bibitem{herzenberg58}
A.~{Herzenberg}, Royal Society of London Philosophical Transactions Series A
  \textbf{250}, 543 (1958)

\bibitem{roberts70}
G.O. {Roberts}, Royal Society of London Philosophical Transactions Series A
  \textbf{266}, 535 (1970)

\bibitem{paxton11}
B.~{Paxton}, L.~{Bildsten}, A.~{Dotter}, F.~{Herwig}, P.~{Lesaffre},
  F.~{Timmes}, ApJS \textbf{192}, 3 (2011)

\bibitem{brun04}
A.S. {Brun}, M.S. {Miesch}, J.~{Toomre}, ApJ \textbf{614}, 1073 (2004)

\bibitem{browning06}
M.K. {Browning}, M.S. {Miesch}, A.S. {Brun}, J.~{Toomre}, ApJL \textbf{648},
  L157 (2006)

\bibitem{brown10}
B.P. {Brown}, M.K. {Browning}, A.S. {Brun}, M.S. {Miesch}, J.~{Toomre}, ApJ
  \textbf{711}, 424 (2010)

\bibitem{racine11}
{\'E}.~{Racine}, P.~{Charbonneau}, M.~{Ghizaru}, A.~{Bouchat}, P.K.
  {Smolarkiewicz}, ApJ \textbf{735}, 46 (2011)

\bibitem{brown11a}
B.P. {Brown}, M.S. {Miesch}, M.K. {Browning}, A.S. {Brun}, J.~{Toomre}, ApJ
  \textbf{731}, 69 (2011)

\bibitem{nelson13a}
N.J. {Nelson}, B.P. {Brown}, A.S. {Brun}, M.S. {Miesch}, J.~{Toomre}, ApJ
  \textbf{762}, 73 (2013)

\bibitem{kapyla13}
P.J. {K{\"a}pyl{\"a}}, M.J. {Mantere}, E.~{Cole}, J.~{Warnecke},
  A.~{Brandenburg}, ArXiv e-prints  (2013)

\bibitem{charbonneau13}
P.~{Charbonneau}, Nature \textbf{493}, 613 (2013)

\bibitem{leighton69}
R.B. {Leighton}, ApJ \textbf{156}, 1 (1969)

\bibitem{glatzmaier84}
G.A. {Glatzmaier}, J. Computational Physics \textbf{55}, 461 (1984)

\bibitem{gilman81}
P.A. {Gilman}, G.A. {Glatzmaier}, ApJS \textbf{45}, 335 (1981)

\bibitem{gilman83}
P.A. {Gilman}, ApJS \textbf{53}, 243 (1983)

\bibitem{brown11b}
B.P. {Brown}, M.K. {Browning}, A.S. {Brun}, M.S. {Miesch}, J.~{Toomre},
  \textbf{448}, 277 (2011)

\bibitem{nelson11}
N.J. {Nelson}, B.P. {Brown}, A.S. {Brun}, M.S. {Miesch}, J.~{Toomre}, ApJL
  \textbf{739}, L38 (2011)

\bibitem{ghizaru10}
M.~{Ghizaru}, P.~{Charbonneau}, P.K. {Smolarkiewicz}, ApJL \textbf{715}, L133
  (2010)

\bibitem{gastine12}
T.~{Gastine}, L.~{Duarte}, J.~{Wicht}, A\&A \textbf{546}, A19 (2012)


\bibitem{nelson13b}
N.J. {Nelson}, B.P. {Brown}, A.~{Sacha Brun}, M.S. {Miesch}, J.~{Toomre}, Solar
  Physics p.~20 (2013)

\bibitem{duarte16}
L.D.V. {Duarte}, J.~{Wicht}, M.K. {Browning}, T.~{Gastine}, MNRAS \textbf{456},
  1708 (2016)

\bibitem{kapyla12}
P.J. {K{\"a}pyl{\"a}}, M.J. {Mantere}, A.~{Brandenburg}, ApJL \textbf{755}, L22
  (2012)

\bibitem{augustson15}
K.~{Augustson}, A.S. {Brun}, M.~{Miesch}, J.~{Toomre}, ApJ \textbf{809}, 149
  (2015)

\bibitem{miesch16}
M.S. {Miesch}, M.~{Zhang}, K.C. {Augustson}, ApJL \textbf{824}, L15 (2016)

\bibitem{brandenburg05}
A.~{Brandenburg}, K.~{Subramanian}, Physics Reports \textbf{417}, 1 (2005)

\bibitem{blackman03}
E.G. {Blackman}, A.~{Brandenburg}, ApJL \textbf{584}, L99 (2003)

\bibitem{warnecke12}
J.~{Warnecke}, P.J. {K{\"a}pyl{\"a}}, M.J. {Mantere}, A.~{Brandenburg}, Solar
  Physics \textbf{280}, 299 (2012)

\bibitem{warnecke13a}
J.~{Warnecke}, P.J. {K{\"a}pyl{\"a}}, M.J. {Mantere}, A.~{Brandenburg}, ArXiv
  e-prints  (2013)

\bibitem{warnecke13b}
J.~{Warnecke}, I.R. {Losada}, A.~{Brandenburg}, N.~{Kleeorin},
  I.~{Rogachevskii}, ArXiv e-prints  (2013)

\bibitem{guerrero16}
G.~{Guerrero}, P.K. {Smolarkiewicz}, E.M. {de Gouveia Dal Pino}, A.G.
  {Kosovichev}, N.N. {Mansour}, ApJ \textbf{819}, 104 (2016)

\bibitem{augustson12}
K.C. {Augustson}, B.P. {Brown}, A.S. {Brun}, M.S. {Miesch}, J.~{Toomre}, ApJ
  \textbf{756}, 169 (2012)

\bibitem{yadav16}
R.~{Yadav}, T.~{Gastine}, U.~{Christensen}, S.J. {Wolk}, K.~{Poppenhaeger},
  Proc. Nat. Acad. Sci. \textbf{113} (2016)

\bibitem{christensen10}
U.R. {Christensen}, Space Science Reviews \textbf{152}, 565 (2010)

\bibitem{augustson17}
K.~{Augustson}, S.~{Mathis}, A.S. {Brun}, 19th Cool Stars Proceedings (2017)

\bibitem{augustson16}
K.C. {Augustson}, A.S. {Brun}, J.~{Toomre}, ApJ \textbf{829}, 92 (2016)

\bibitem{roberts13}
P.H. Roberts, E.M. King, Reports on Progress in Physics \textbf{76}, 096801
  (2013)

\bibitem{davidson13}
P.A. {Davidson}, Geophys. J. International \textbf{195}, 67 (2013)

\bibitem{christensen09}
U.R. {Christensen}, V.~{Holzwarth}, A.~{Reiners}, Nature \textbf{457}, 167
  (2009)
\end{thebibliography}
\end{document}